\documentclass[sigconf]{acmart}

\usepackage{balance}
\usepackage{siunitx}
\usepackage[caption=false,font=footnotesize,farskip=0pt]{subfig}%
\usepackage{amsmath}

\usepackage[utf8]{inputenc}

\AtBeginDocument{%
  \providecommand\BibTeX{{%
    \normalfont B\kern-0.5em{\scshape i\kern-0.25em b}\kern-0.8em\TeX}}}

\copyrightyear{2021}
\acmYear{2021}
\setcopyright{acmcopyright}\acmConference[NOSSDAV '21]{Workshop on Network and Operating System Support for Digital Audio and Video (NOSSDAV '21))}{September 28-October 1, 2021}{Istanbul, Turkey}
\acmBooktitle{Workshop on Network and Operating System Support for Digital Audio and Video (NOSSDAV '21)) (NOSSDAV '21), September 28-October 1, 2021, Istanbul, Turkey}
\acmPrice{15.00}
\acmDOI{10.1145/3458306.3461000}
\acmISBN{978-1-4503-8435-3/21/09}

\setcopyright{none}
\renewcommand\footnotetextcopyrightpermission[1]{} 

\begin{document}

\title{Viewport-Aware Dynamic \ang{360} Video Segment Categorization}



\author{Amaya Dharmasiri}
\email{minikiraniamaya@gmail.com}
\affiliation{%
  \institution{The University of Sydney}
}
\author{Chamara Kattadige}
\email{ckat9988@uni.sydney.edu.au}
\affiliation{%
  \institution{The University of Sydney}
}

\author{Vincent Zhang}
\email{vzha9726@uni.sydney.edu.au}
\affiliation{%
  \institution{The University of Sydney}
}

\author{Kanchana Thilakarathna}
\email{kanchana.thilakarathna@sydney.edu.au}
\affiliation{%
  \institution{The University of Sydney}
}

\renewcommand{\shortauthors}{Dharmasiri, et al.}

\begin{abstract}

Unlike conventional videos, \ang{360} videos give freedom to users to turn their heads, watch and interact with the content owing to its immersive spherical environment. Although these movements are arbitrary, similarities can be observed between viewport patterns of different users and different videos. 
Identifying such patterns can assist both content and network providers to enhance the \ang{360} video streaming process, eventually increasing the end-user Quality of Experience (QoE). 
But a study on how \textit{viewport patterns display similarities across different video content, and their potential applications} has not yet been done.
In this paper, we present a comprehensive analysis of a dataset of 88 \ang{360} videos and propose a novel video categorization algorithm that is based on similarities of viewports.
First, we propose a novel viewport clustering algorithm that outperforms the existing algorithms in terms of clustering viewports with similar positioning and speed.
Next, we develop a novel and unique dynamic video segment categorization algorithm that shows 
notable improvement in similarity for viewport distributions within the clusters when compared to that of existing static video categorizations. 



\end{abstract}

%
%





\maketitle

\section{Introduction}

With recent advances of vision technologies on smartphones and Head Mounted Devices such as Facebook Oculus~\cite{fb_oculus}, Samsung Gear VR~\cite{samsung_gear}, Microsoft Hololens~\cite{MS_hololens}, VR/\ang{360} video streaming is becoming increasingly popular. The major service providers such as YouTube and Facebook have already started streaming \ang{360} videos through optimized spherical video players~\cite{fb_360,yt_360}. 


\ang{360} video frames are typically 4-6 times larger than normal videos ~\cite{qian2018flare} and it can grow up to 80 times~\cite{guan_pano_2019}, requiring more resources at every entity involved in the process, i.e. servers, networks and end-user devices. Viewport-adaptive streaming, \textit{a.k.a.} Field-of-View (FoV) aware streaming, has emerged as an efficient approach for \ang{360} video streaming \cite{qian_optimizing_2016, qian2018flare,he2018rubiks}. FoV-aware streaming is expected to significantly reduce the amount of data that has to be transferred by predicting the next viewport or FoV (visible area) of the user and sending only the selected portion of the panoramic frame to the client. Despite the great promise and research, viewport-adaptive streaming has not yet been adopted by streaming service providers like YouTube and Facebook. This is primarily due to lack of end-device software support in viewport-adaptive streaming and the lack of holistic understanding of how user viewports change from frame to frame and from video to video.

There have been scattered efforts to understand viewport characteristics of \ang{360} videos, but with small-scale experiments with limited number of videos~\cite{xie_cls_2018,rossi_spherical_2019,petrangeli_trajectory-based_2018}. This paper presents a comprehensive analysis of viewport trajectories of \ang{360} video users by analysing a large unified dataset of over 3700 viewport traces. \emph{To the best of our knowledge, this is the largest \ang{360} video user behavior analysis conducted to date.} We propose a viewport clustering mechanism to understand similarities of user behaviors that takes spatial, motion and other behavioral patterns into account. We particularly highlight the importance of adding temporal features such as head movement speed to the viewport clustering to ensure that subtle differences in highly dynamic viewport patterns can be captured. 

Video categorizations are widely studied for normal video streaming~\cite{sun2016exploiting,jiang2017exploiting}, which 
can support video indexing, storage management, popularity prediction, and efficient resource provisioning at both content and network service providers~\cite{sun2016exploiting, mcclanahan2017interplay}. 
However, these traditional video categorizations will not be effective anymore for \ang{360} videos as they categorize entire video into one category or cluster~\cite{afzal_characterization_2017, carlsson_had_2020, nasrabadi_taxonomy_2019}. Due to the freedom of users to request different portions of the same frame and tile-based streaming of \ang{360} videos, user behavior can vary significantly even within a given video. In this paper, we propose a novel video segment categorization mechanism based on the dynamic viewport characteristics of users that cluster chunks of the same video into different groups. 

The key contributions of this paper are;\vspace{-1mm}
\begin{itemize}
    \item An aggregated dataset with 88 videos with over 3700 viewport traces containing over 142 hours of viewing logs. We consolidate different formats of individual datasets, e.g. head orientation, data sampling rates, trace duration, into unified framework, which has been made publicly available\footnote{\url{https://github.com/theamaya/Viewport-Aware-Dynamic-360-Video-Segment-Categorization}}.
    \item A novel feature based viewport clustering algorithm that takes spatial, motional and other behavioral features of viewport patterns into account. We validate the effectiveness of the proposed clustering mechanism using the aggregated dataset compared to previously proposed clustering methods. The results show an average of 81.17\% of viewport overlap and limits the pairwise difference in angular speed of 0.47 rad/s for the traces within the resulting clusters.
    \item A novel dynamic video segment categorization algorithm which is based on different user behavior types induced by the corresponding video. The results show 28.32\% improvement in similarity in viewport distribution within the clusters when compared to the existing static genre-based video categorizations. 
\end{itemize}

\section{Related Work}

\textbf{Viewport clustering}. 
Majority of the existing algorithms for viewport clustering, are based on the hypothesis that users tend to exhibit similar viewing patterns for a given video. One notable clustering approach, proposed by \cite{rossi_spherical_2019} leverages the spatial arrangement of user viewpoints at a particular play-point or during a specific time window of the video. The clustering is done in the spherical domain using geodesic distance as the closeness metric. Another approach proposed by \cite{xie_cls_2018} extracts user fixations from viewport traces and performs density-based clustering using a spatial distance metric. The algorithm proposed by \cite{petrangeli_trajectory-based_2018} utilizes viewport trajectory data in the long term and clusters the viewports with a time and distance based metric. 

\textbf{Video categorization}. 
The literature surrounding video categorization is limited, with most studies focusing on video titles and genres. Afzal \textit{et al.}~\cite{afzal_characterization_2017} categorized 2285 \ang{360} videos based on their genre which is determined by keywords and found 14 different categories, ranging from documentaries to sports and driving. Another categorization approach, proposed by \cite{almquist_prefetch_2018} and \cite{carlsson_had_2020} used manual inspection of video content and expected behaviors of user FoV trajectories, resulting in categories such as \textit{Exploration}, \textit{Static focus}, \textit{Rides} etc. An approach proposed by \cite{nasrabadi_taxonomy_2019} categorizes videos based on the extent of camera motion and the number of moving targets. 

\textit{In contrast to these methods, we aim to leverage a viewport-aware approach to the problem of \ang{360} video categorization. Our proposed solution aims to go beyond spatial features by considering temporal and behavioral patterns as metrics for viewport similarity modeling.}

\begin{table*}[t]
  \small
  \caption{Summary of Datasets }\vspace{-4mm}
  \label{tab:dataset}
  \begin{tabular}{p{1cm} p{0.7cm} p{5cm} p{1.3cm} p{3.2cm} p{1.2cm} p{2.0cm} }
    \toprule
    Dataset  & \# of videos & Genres/ Categories & \# of traces per video & Viewport trace representation (Quat: Quaternion) & Sampling rate & Video duration\\
    \midrule
    \text{1-~\cite{corbillon_360-degree_2017}} & 7 & Documentary, Landscapes, Entertainment & 50 & Unit Quat& \textasciitilde 40Hz & 1m \\
    \text{2-~\cite{lo_360_2017}} & 10 & Natural image and Computer generated, Fast and slow paced & 50 & Yaw, pitch, roll & 30Hz & 1m \\
    \text{3-~\cite{bao_shooting_2016}} & 16 & Landscapes, Entertainment, Sports & 60 & Yaw, pitch, roll & \textasciitilde 7-9Hz & 30s \\
    \text{4-~\cite{wu_dataset_2017}} & 18 & Documentary, Performance, Film, Sports & 48 & Unit Quat & \textasciitilde 9Hz & 2m 52s- 10m 55s \\
    \text{5-~\cite{guan_pano_2019}} & 18 & Documentary, Sports, Performance, Other & 48 & Yaw, pitch & 30Hz & 55s - 3m 40s \\
    \text{6-~\cite{nasrabadi_taxonomy_2019}} & 28 & 15 categories based on camera motion and no. of moving targets & 30 & Unit Quat, Spherical vectors & 60Hz & 1m \\
    \bottomrule
  \end{tabular}
\end{table*}

\vspace{-2mm}
\section{Dataset}\label{sec:dataset}
We created a unified dataset  combining 6 different \ang{360} video datasets from the literature. The selected datasets contain \ang{360} videos 
and their corresponding user head orientation data. Table~\ref{tab:dataset} summarises the
datasets used. The aggregated dataset contains 88 videos with an average of 45 user traces per video. Each video is at least 30s long. We have made the aggregated dataset publicly available \href{https://github.com/theamaya/Viewport-Aware-Dynamic-360-Video-Segment-Categorization/}{\underline{here}}. 

First, 
we converted all head orientation traces to the same representation; Yaw and Pitch angles in radians (Roll angle variation was negligible, therefore was ignored in this study). Next, 
to tackle the differences in sampling rates, we re-sampled the entire dataset at 10Hz.
For this study, we only considered the datasets collected using
HMDs, excluding
mobile and PC users to eliminate the impact of device on viewing patterns~\cite{xu_analyzing_2019}\footnote{We keep viewport patterns from different rendering devices and its impact on our categorization task as a future work}.



\vspace{-2mm}
\begin{figure}[h]
\centering
    \subfloat[Viewport fixations for 4 sample videos]{\includegraphics[trim=0 0 0 0, clip,width=0.7\columnwidth]{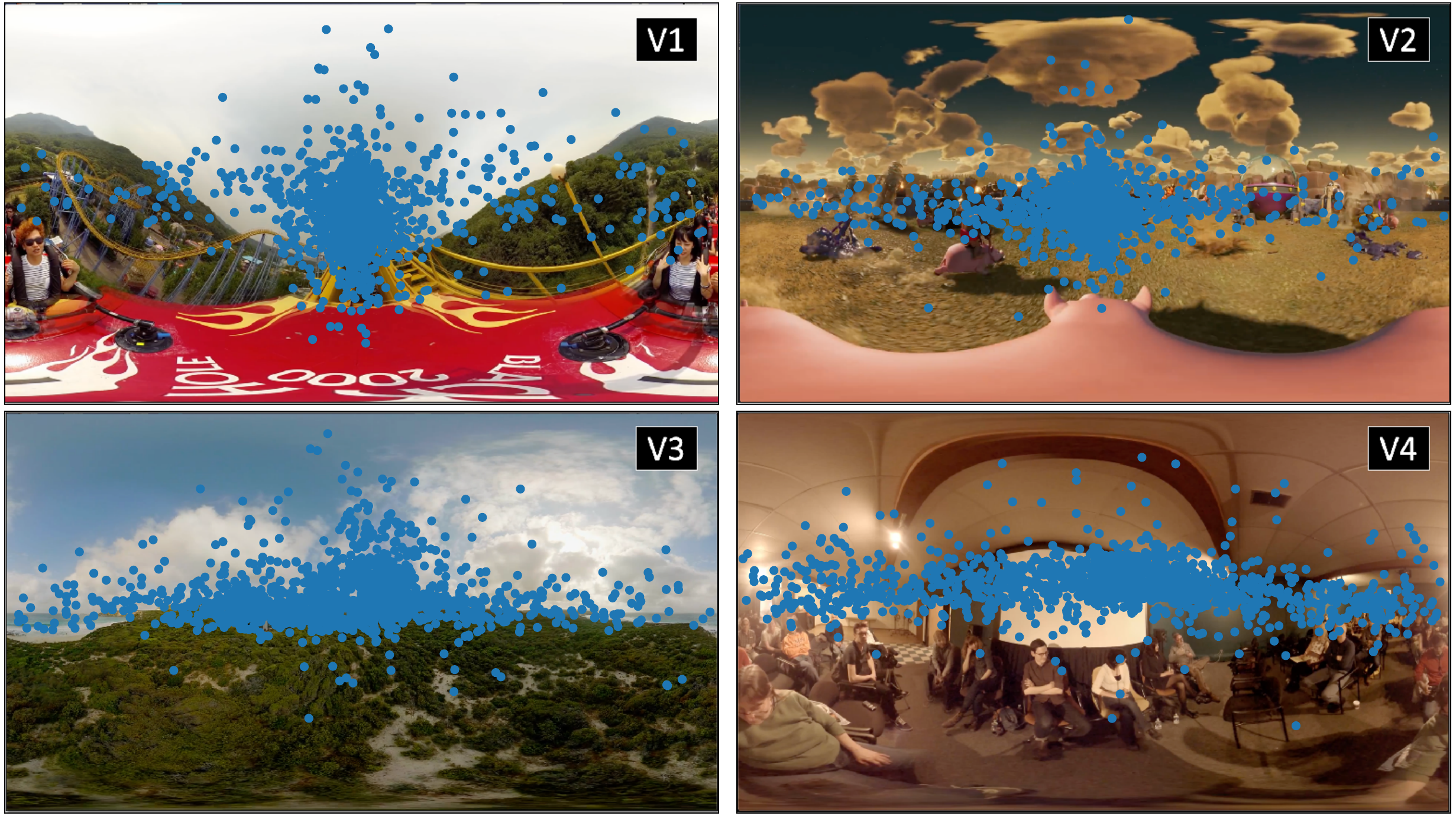}%
    \label{fig:motivation_2_diff_videos}}
    
    \subfloat[Mean distribution: Yaw]{\includegraphics[width=0.35\columnwidth]{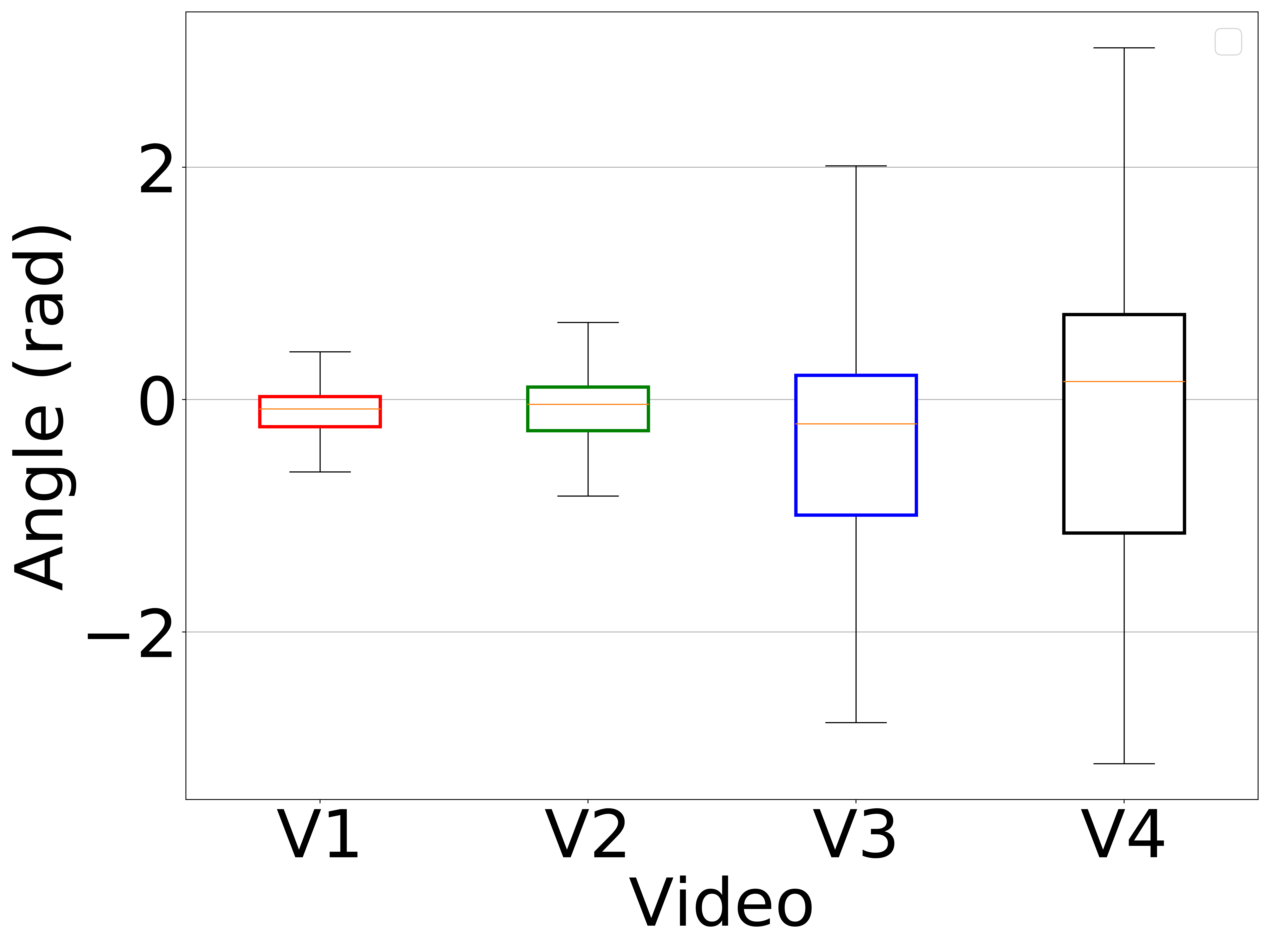}%
    \label{fig:motivation_f3_dist}}\quad
    \hspace{0.5mm}
    \subfloat[Max yaw movement]{\includegraphics[width=0.35\columnwidth]{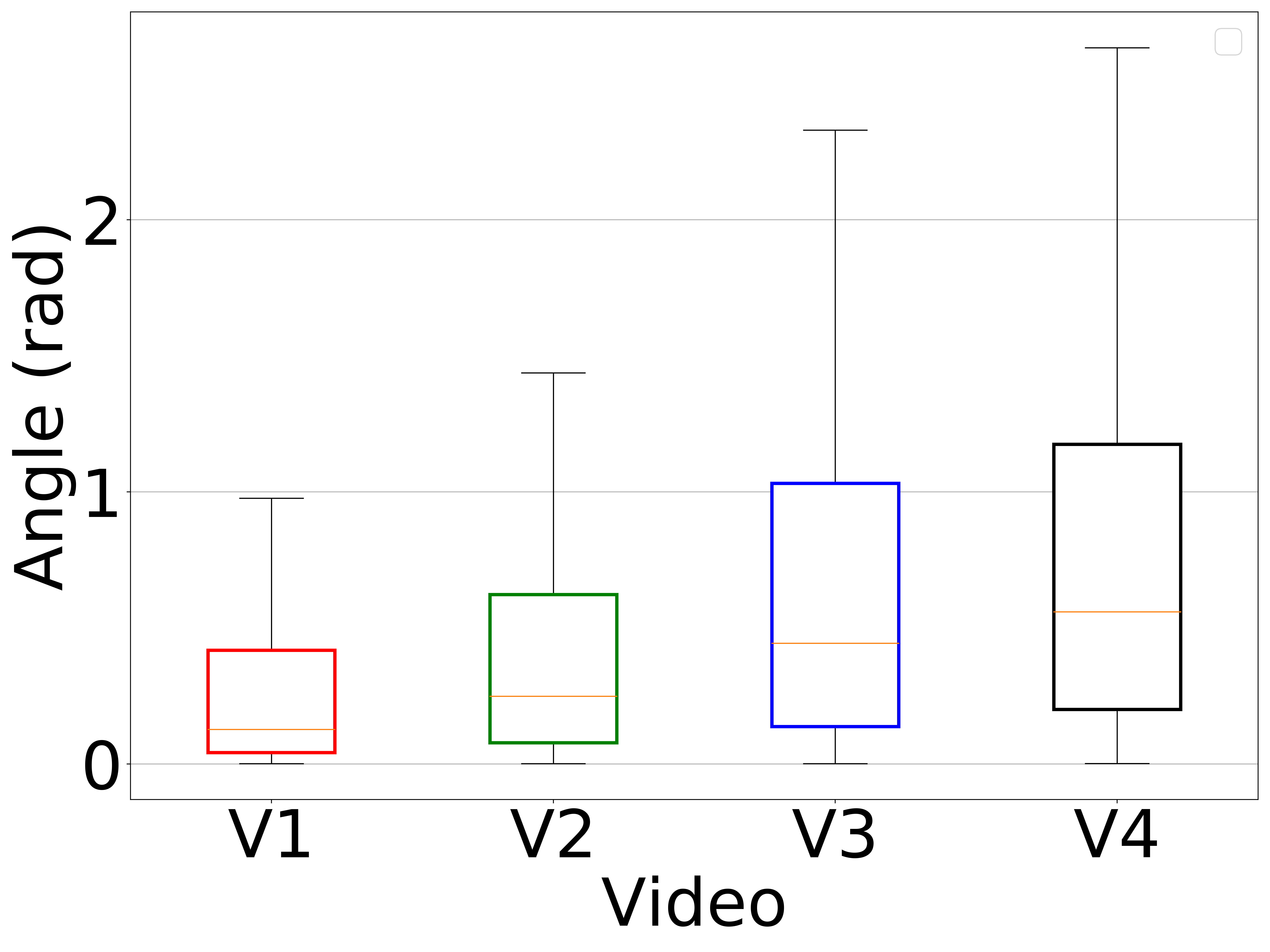}
    \label{fig:motivation_f11_dist}}
    
    \vspace{-3mm}
    \caption{Viewport and two feature distribution between selected videos in two categories}
    \vspace{-4mm}
    \label{fig:motivation}
\end{figure}


\vspace{-2mm}
\subsection{Relationship of viewport and video types}

In our preliminary analysis, we studied how viewport distributions differ between different videos. Fig.~\ref{fig:motivation} shows four example videos analyzed in this study. In both \textit{i)} V1 (Rollercoster) \& V2 (Riding on a Hog): the camera is moving forward with a high speed whereas in \textit{ii)} V3 (Landscape) \& V4 (Panel): Camera movement is not significant, and contents are dispersed in the frame.

Viewport distributions in Fig.~\ref{fig:motivation_2_diff_videos} shows that, content of the video has an impact on the users' viewport distribution. We see that the viewports of V1 \& V2 are centered in the frame whereas, viewports are comparably dispersed in V3 \& V4. This indicates that, 
viewport distributions can be used as a proxy to explain certain content patterns in \ang{360} videos
indicating that a viewport based categorization of videos can be effective. Fig.~\ref{fig:motivation_f3_dist} plots the spatial distribution of 2s viewport chunks in the yaw axis, and~\ref{fig:motivation_f11_dist} plots the maximum yaw angle moved by a viewport trace within a 2s chunk. We observe that patterns of raw viewport distributions in \ref{fig:motivation_2_diff_videos} are reflected in these features, indicating that carefully extracted features from viewport traces can result in an effective video categorization.




\vspace{-1mm}
\section{feature-based viewport clustering}\label{section 4}

In this section, we propose a feature representation for user viewport traces, which is used for a novel viewport clustering method. In Section \ref{section 5}, we extend this feature based viewport clustering to develop a novel dynamic video categorization. 
We divided the duration of a video/ viewport trace into bins (chunks) of 2 seconds. 
Table~\ref{tab:basic notation} briefly describes the notations used in our further explanations.






\begin{table}[h]
  \small
  \vspace{-2mm}
  \caption{Feature selection for viewport clustering}
  \vspace{-4mm}
  \begin{tabular}{l|l} 
    \toprule
    Notation & Description\\
    \midrule
    $i$, $k$ &  video id: $i \in [ 0,87]$, chunk id: $k \in [ 0,13]$,  \\
    $n_i$ & num. of users for $i^{th}$ video\\
    $j$ & user id, $j\in [ 0,n_{j})$\\
    $\{V_{i}\}$ & Set of all videos $\forall i \in [ 0,87]$ \\
    $\{c_{k}\}$ & Set of all time chunks $\forall k \in [ 0,13]$ \\
    $\{u_{i,j}\}$ & Set of all users $\forall i \in [ 0,87], \forall j\in [ 0,n_{i})$ \\
    $\{v_{i,k}\}$ & Set of all video chunks $\forall i \in [ 0,87], \forall k \in [ 0,13]$\\
    $\{t_{i,k,j}\}$ & Set of all viewport trace chunks  $\forall i \in [ 0,87], \forall k \in [ 0,13],$\\
    & $\forall j\in [ 0,n_{i})$\\
    \bottomrule
\end{tabular}
\label{tab:basic notation}
\vspace{-2mm}
\end{table}

\vspace{-2mm}
\subsection{Feature extraction}\label{Section 4- feature extraction}

We observed that the
\textit{users who consume \ang{360} video content differ to each other in multiple aspects such as positioning, speed of head movement, extent of the sphere explored and the maximum angle moved within a defined window}. 



We extracted features for every viewport trace chunk in $\{t_{i,k,j}\}$ as summarized in Table \ref{tab:Feature_selection_for_viewport_clustering}. Let $F^{( 1)}$ represent the feature extraction function for a viewport trace chunk. $\underline{f}^{( 1)}_{i,k,j}$ is a 15 dimensional feature vector that represent a viewport trace chunk in the dataset.

\vspace{-2mm}
\begin{equation}\label{eq:feature_extraction_func_vp_clust}
 F^{( 1)}( t_{i,k,j}) =\underline{f}^{( 1)}_{i,k,j}
\end{equation}
\vspace{-3mm}

\begin{table}[h]
  \small
  \caption{Feature selection for viewport clustering}
  \vspace{-4mm}
  \begin{tabular}{p{2.0cm}| p{5.5cm}} 
    \toprule
    Feature & Measurements y--yaw, p--pitch\\
    \midrule
    Position & mean, 25\textsuperscript{th} and 75\textsuperscript{th} percentiles: y \& p \\
    Speed &  mean, 25\textsuperscript{th} and 75\textsuperscript{th} percentiles: y \& p \\
    Maximum angle & 
    angular position relative to initial point: y \& p\\
    Area explored & \% coverage on the sphere by \ang{100}x\ang{100}  viewport\\
  \bottomrule
\end{tabular}
\label{tab:Feature_selection_for_viewport_clustering}
\end{table}

\vspace{-2mm}
\subsection{Clustering the viewport trace chunks}

\begin{figure}[h]
\centering
\includegraphics[width=0.9\columnwidth]{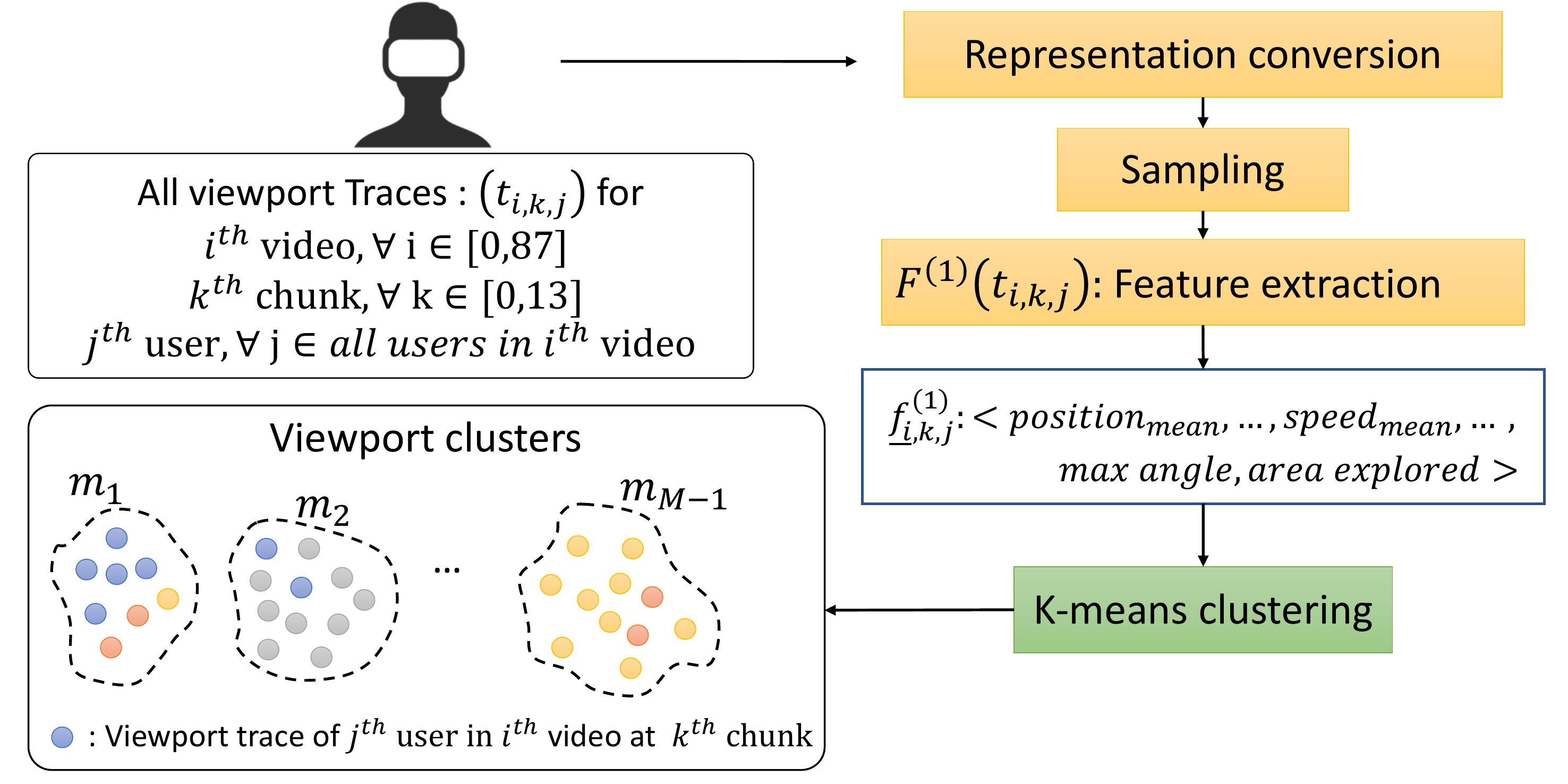}%
\label{fig:viewport_clustering_based_categorization}
\vspace{-4mm}
\caption{Overview of viewport clustering}
\label{fig:view_port_based_cat}
\vspace{-4mm}
\end{figure}

Fig.~\ref{fig:view_port_based_cat} summarizes the complete procedure of generating feature vector representations for viewport traces
which are then used as input to  a K-Means algorithm that clusters the viewport trace chunks into \textit{M} clusters. 
We extract the cluster to which a viewport trace chunk belongs to using Eq.~\ref{eq:cluster_alloction_func} (further used in Section \ref{section 5})


\begin{equation}\label{eq:cluster_alloction_func}
    C^{( 1)}( t_{i,k,j}) =m,\ m\in [ 0,M)
\end{equation}




We defined three pairwise metrics to quantify the similarity of two viewport trace chunks for this viewport cluster analysis.
\begin{itemize}
\item\textit{Pairwise viewport overlap- VPO-} \cite{rossi_spherical_2019} measures the spatial coherence of two traces. It calculates the geodesic distance between each two corresponding sample points of 2 viewport traces and uses it as a proxy for pairwise viewport overlap. In our representation, a viewport trace chunk has 20 points (10Hz, 2 seconds). 
The average of geodesic distance between each pair of corresponding points is taken as pairwise VPO.

\item\textit{Pairwise difference in head movement speed-}The average angular speed is calculated for each viewport trace segment, and their pairwise absolute difference is measured. 

\item\textit{Pairwise difference in percentage of sphere explored-}within the 2s duration from the features extracted as mentioned in \ref{tab:Feature_selection_for_viewport_clustering}. 

\end{itemize}
\textit{maximization} of the Pairwise VPO, and \textit{minimization} of Pairwise difference in head movement speed and Pairwise difference in percentage of sphere explored are desirable within a cluster.

\vspace{-1mm}

\subsubsection{The effectiveness of the proposed feature representation}

Clustering similar user viewport traces is often used in viewport prediction for \ang{360} video streaming. In this experiment we prove that the feature representation introduced in Eq.~\ref{eq:feature_extraction_func_vp_clust} can be used to accurately cluster viewports in such an application. 
The feature vectors $\underline{f}^{( 1)}_{i,k,j},\ \forall j\in [0,n_{i}]$ as input to the K-Means clustering to yield \textit{M} different clusters separately for different videos $i$ and different time bins $k$. 
The results were compared with 3 other algorithms,

\textbf{Spherical Clustering\cite{rossi_spherical_2019}}- Models viewport traces as nodes of a unit distance graph. Edges are connected if the two viewport traces have high viewport overlap. Finds maximal cliques as clusters

\textbf{Trajectory based\cite{petrangeli_trajectory-based_2018}}- Represents viewport traces as graph nodes with edges weighted by their spatial similarity. Uses spectral clustering while minimizing the distortion score.

\textbf{DBSCAN\cite{xie_cls_2018}}- Maximizes spatial density of clusters by taking the mean positions of trajectories in yaw, pitch in the 2s interval


\begin{figure}[h]
\vspace{-2mm}
\centering
\subfloat[Pairwise viewport overlap]{\includegraphics[width=0.18\textwidth]{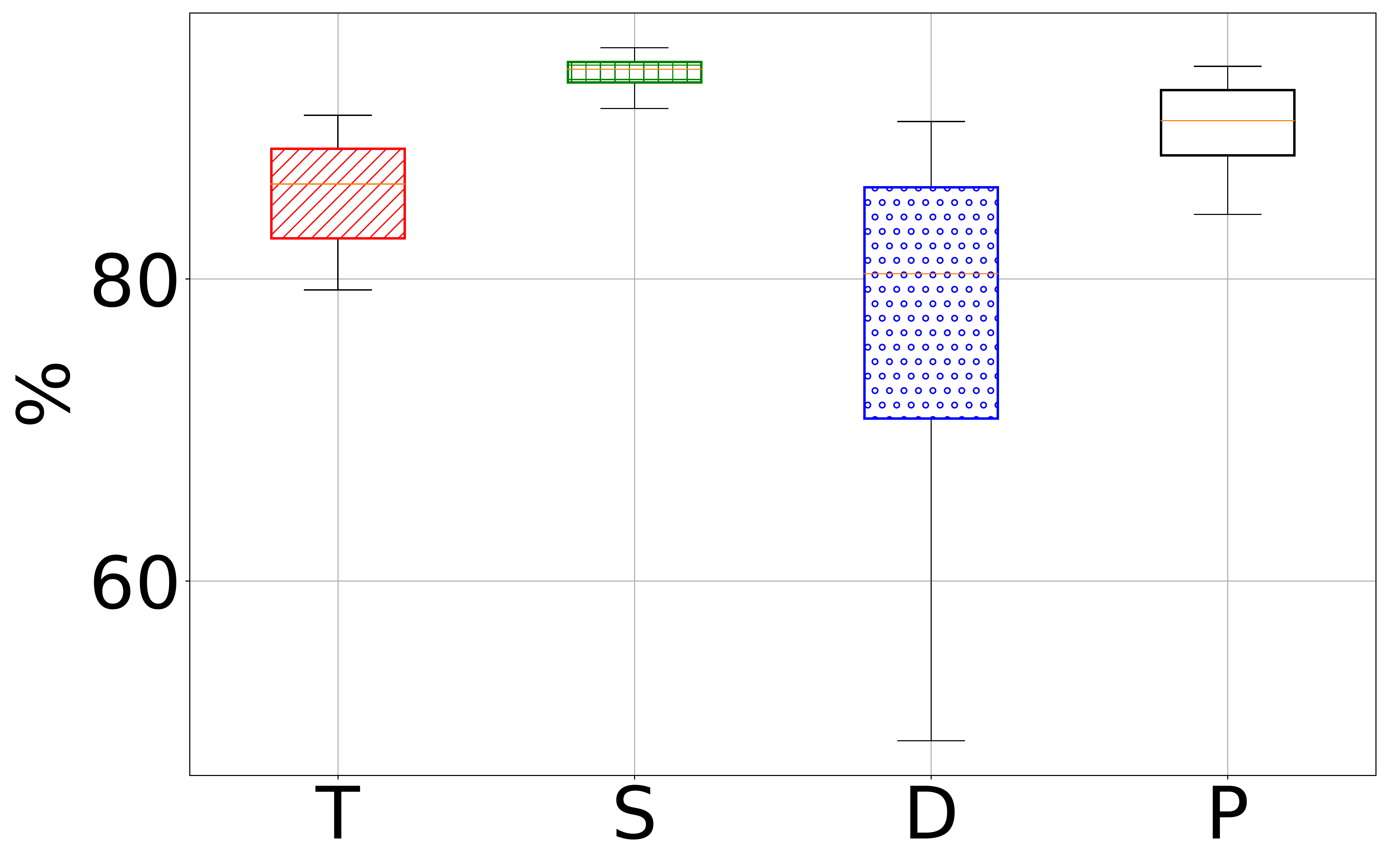}%
    \label{fig:Pairwise viewport overlap}}
    \hspace{2mm}
\subfloat[Pairwise difference in head movement speed]{\includegraphics[width=0.18\textwidth]{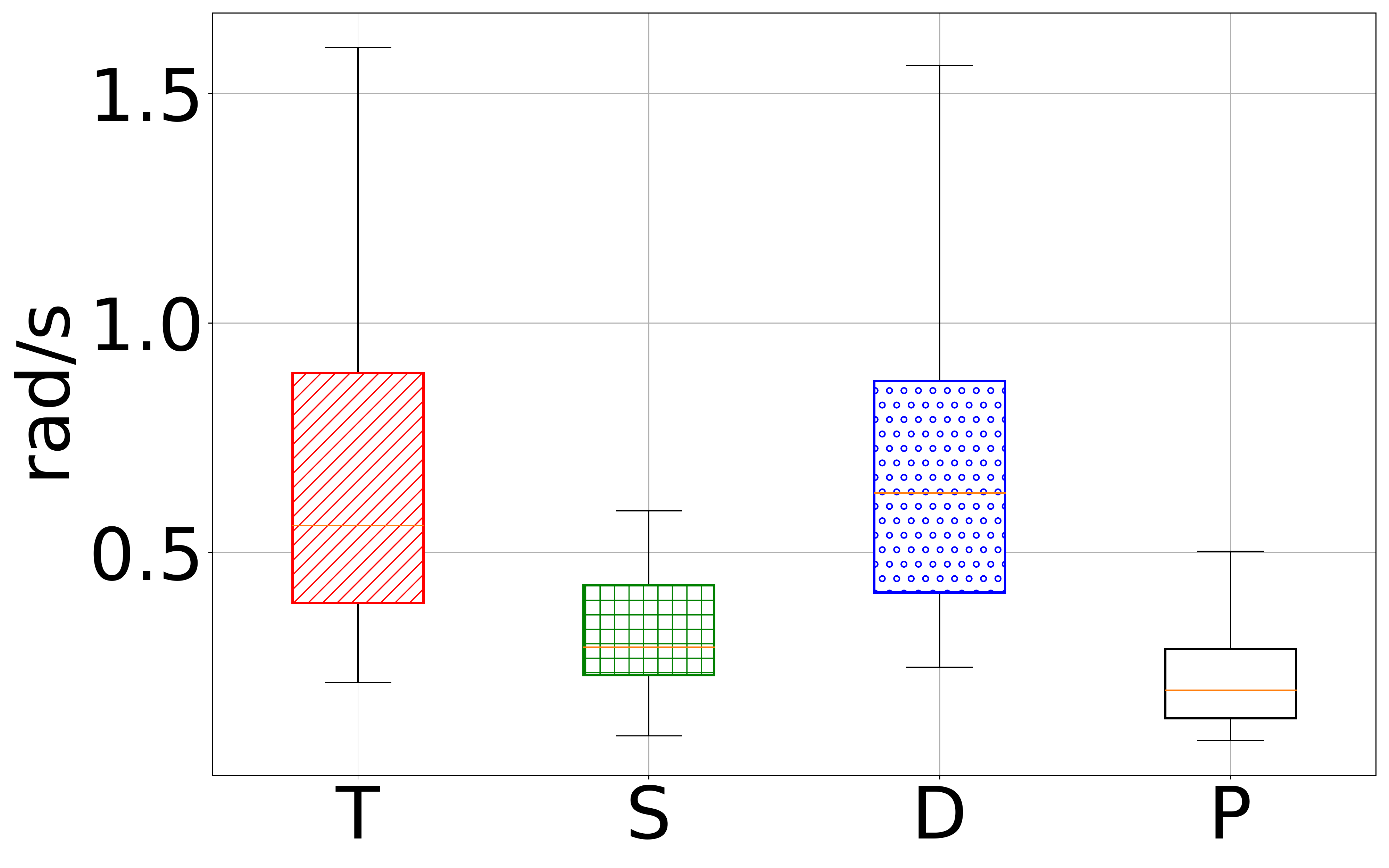}%
    \label{fig:Pairwise difference in head movement speed}}
\vspace{-3mm}
\subfloat[Pairwise difference in percentage of sphere explored ]{\includegraphics[width=0.18\textwidth]{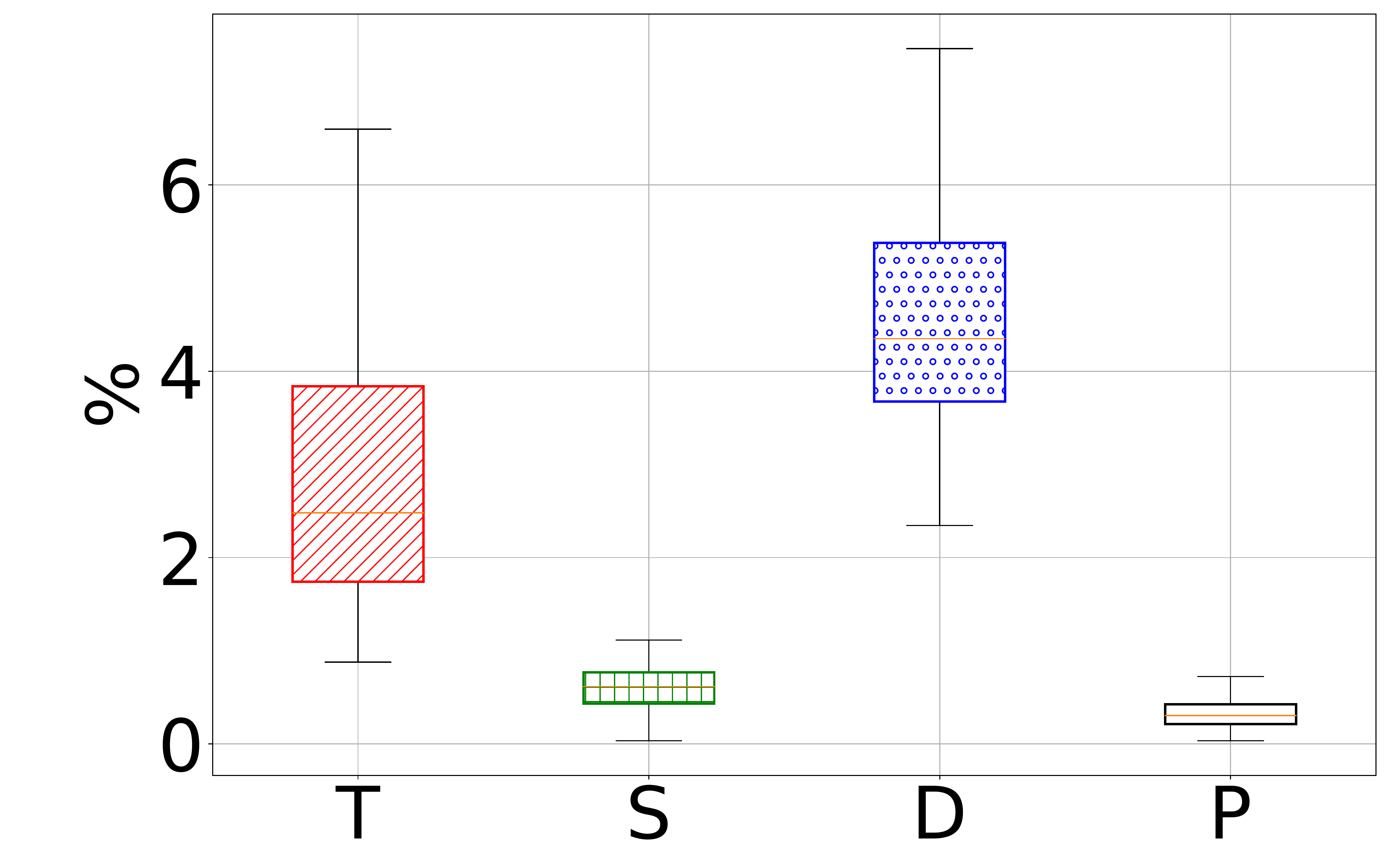}%
    \label{fig:Pairwise difference in percentage of sphere explored }}
    \hspace{2mm}
\subfloat[Number of resulting clusters ]{\includegraphics[width=0.115\textwidth,angle=90 ]{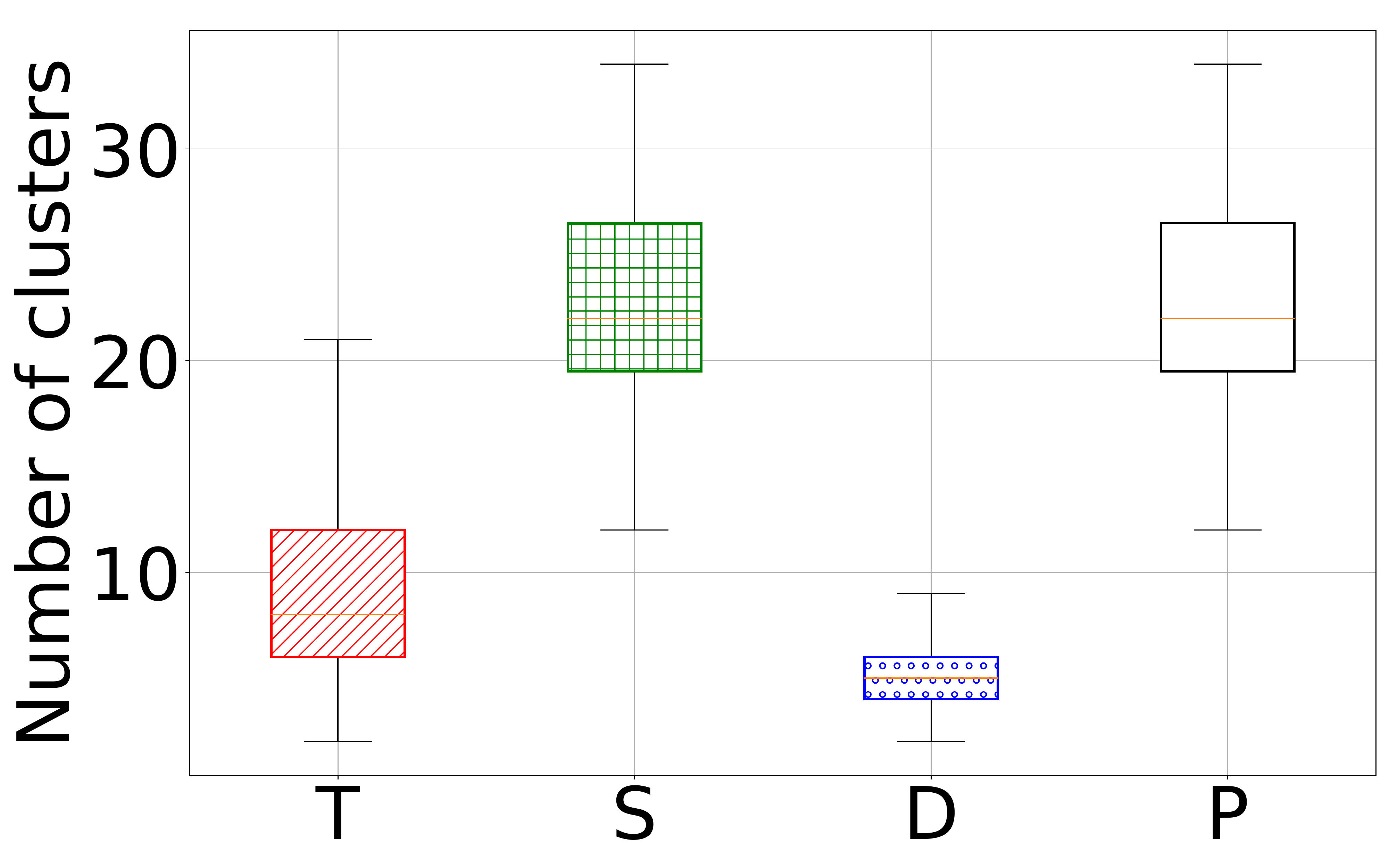}%
    \label{fig:Number of resulting clusters }}
\vspace{-4mm}
\caption{Within-cluster measurement comparison of T-Trajectory based\cite{petrangeli_trajectory-based_2018}, S- Spherical\cite{rossi_spherical_2019}, D- DBSCAN\cite{xie_cls_2018} with the P- proposed for the aggregated dataset.}
\vspace{-2mm}
\label{fig:VP clustering algorithm comparison}
\end{figure}

Fig.~\ref{fig:VP clustering algorithm comparison} contains the aggregated comparison of the proposed algorithm with \cite{rossi_spherical_2019, petrangeli_trajectory-based_2018, xie_cls_2018} for the task of viewport clustering. The analysis was done for all $\{v_{i,k}\}, \forall i \in [ 0,87], \forall k \in [ 0,13]$.
For a clear comparison, at each video and time chunk we used the resulting number of clusters from Spherical clustering\cite{rossi_spherical_2019} as the number of clusters for K-means in the proposed algorithm. It is clear that the proposed algorithm outperforms all others in speed and percentage exploration based metrics, and shows on par performance with Spherical clustering algorithm in \textit{VPO} metric. 

\begin{figure}[h]
\centering
\subfloat[Viewport overlap - rollercoaster]{\includegraphics[width=0.244\textwidth]{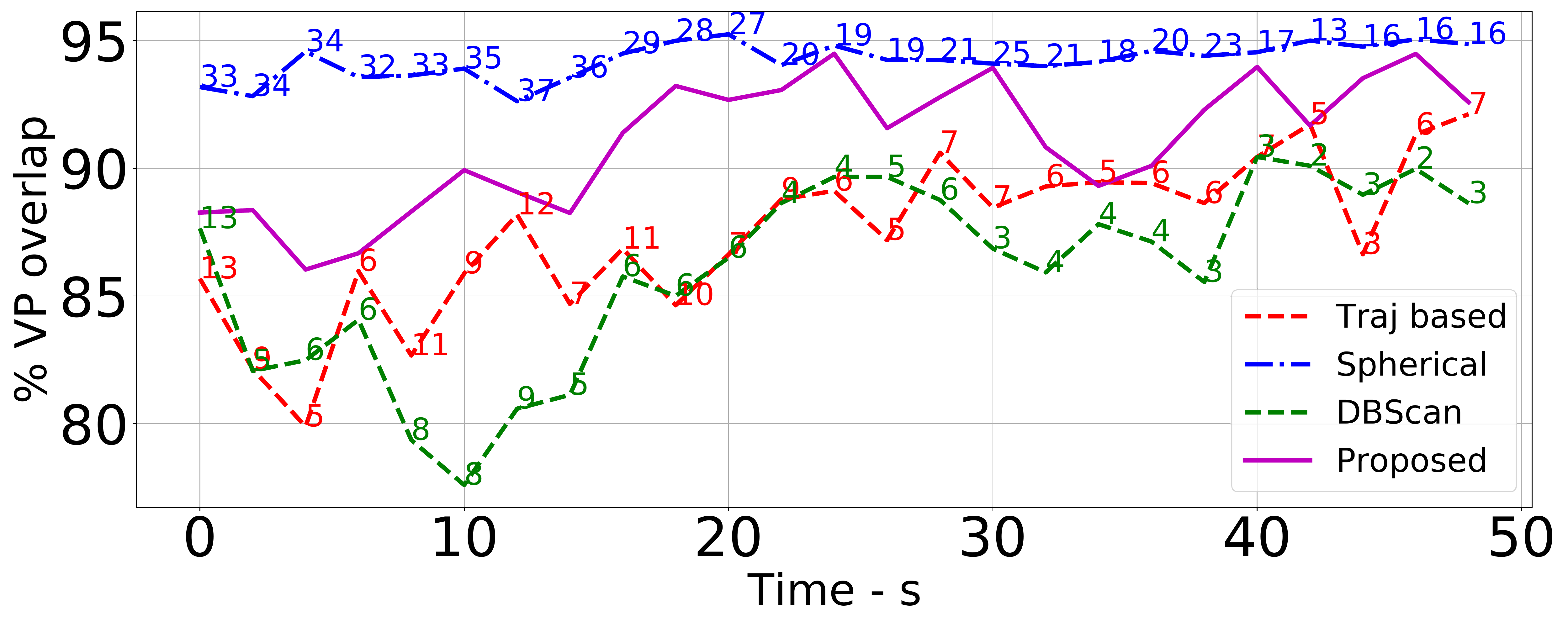}%
    \label{fig:VP overlap rollercoaster}}
\subfloat[Viewport overlap - timelapse]{\includegraphics[width=0.244\textwidth]{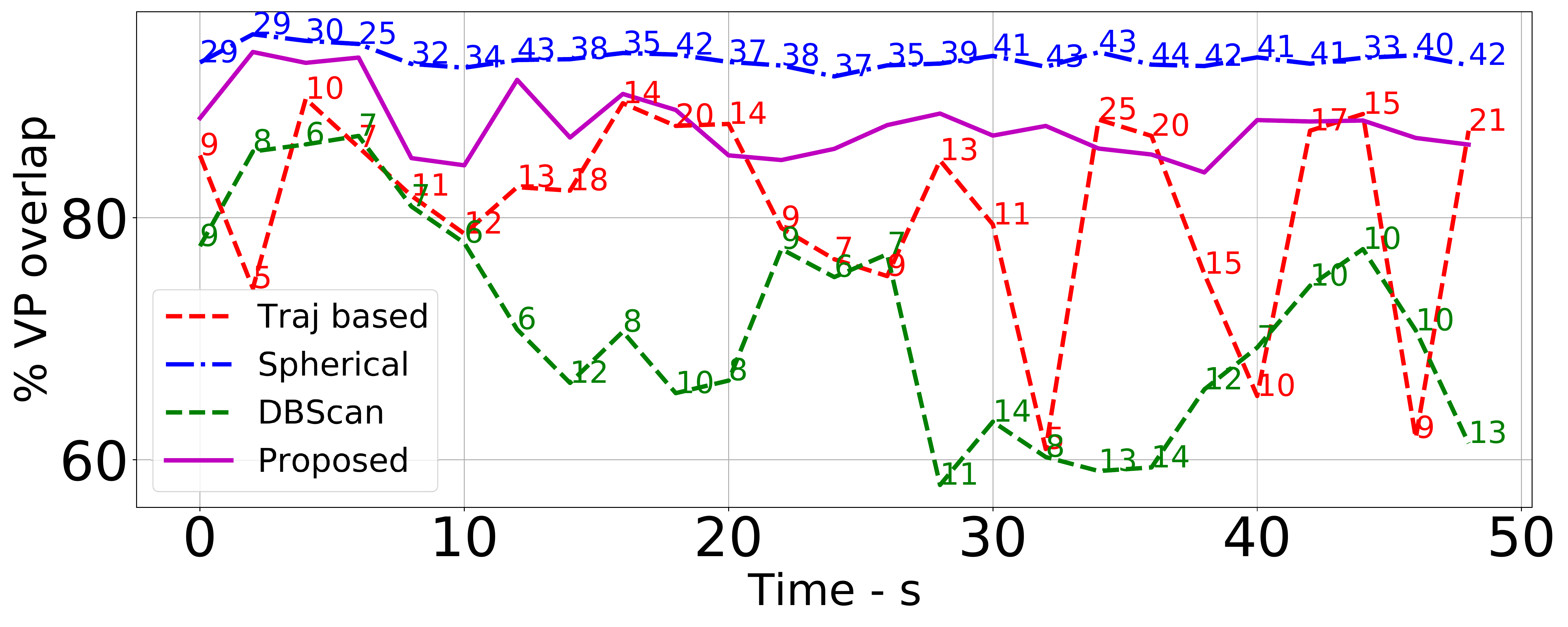}%
    \label{fig:VP overlap timelapse}}
\hfill
\subfloat[Pairwise Speed diff. -rollercoaster ]{\includegraphics[width=0.244\textwidth]{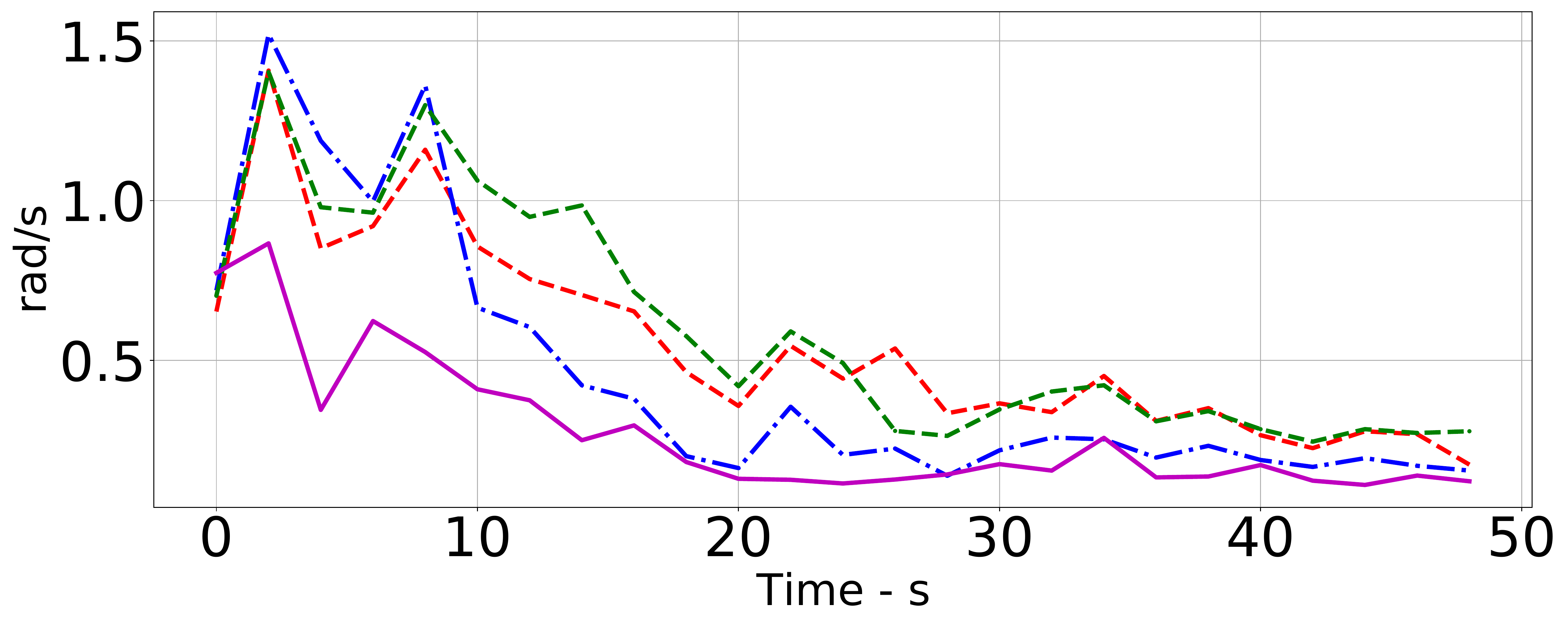}%
    \label{fig:Speed difference rollercoaster}}
\subfloat[Pairwise Speed diff. -timelapse ]{\includegraphics[width=0.244\textwidth]{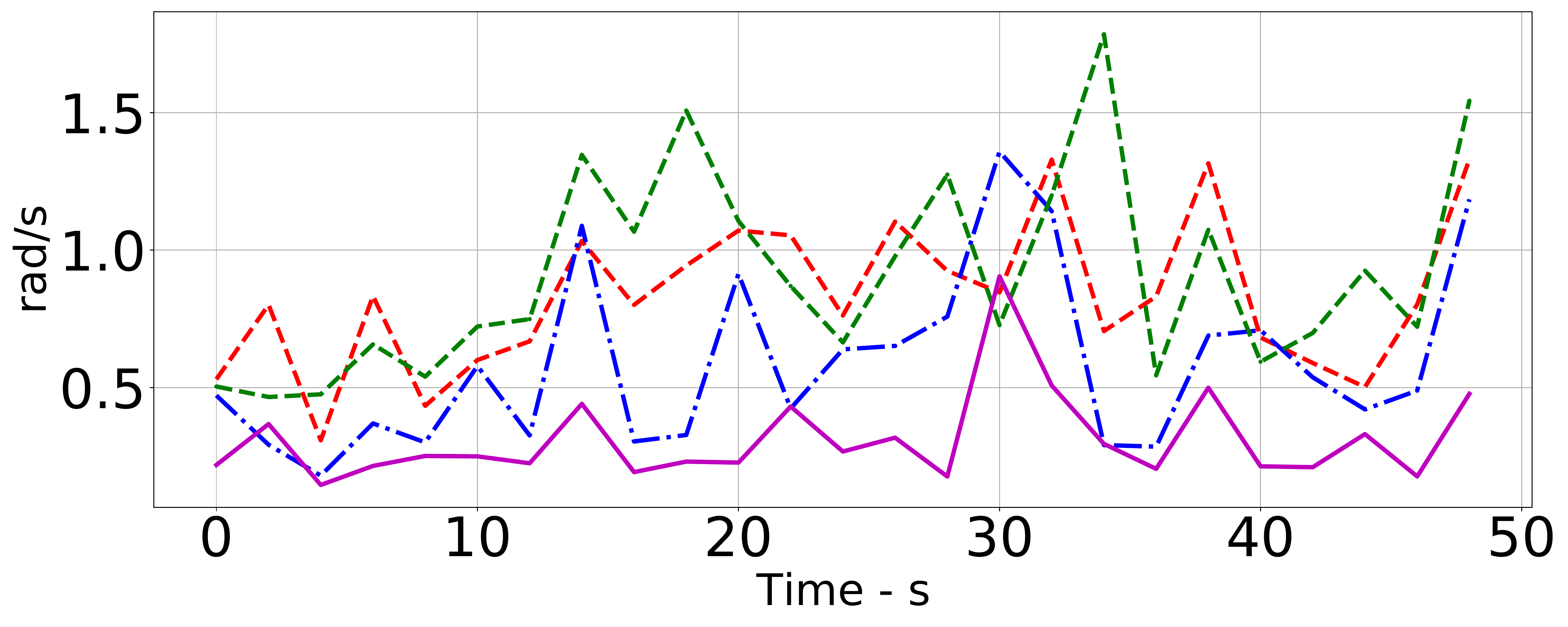}%
    \label{fig:Speed difference timelapse}}
\hfill
\subfloat[Pairwise \% sphere diff. -rollercoaster ]{\includegraphics[width=0.244\textwidth]{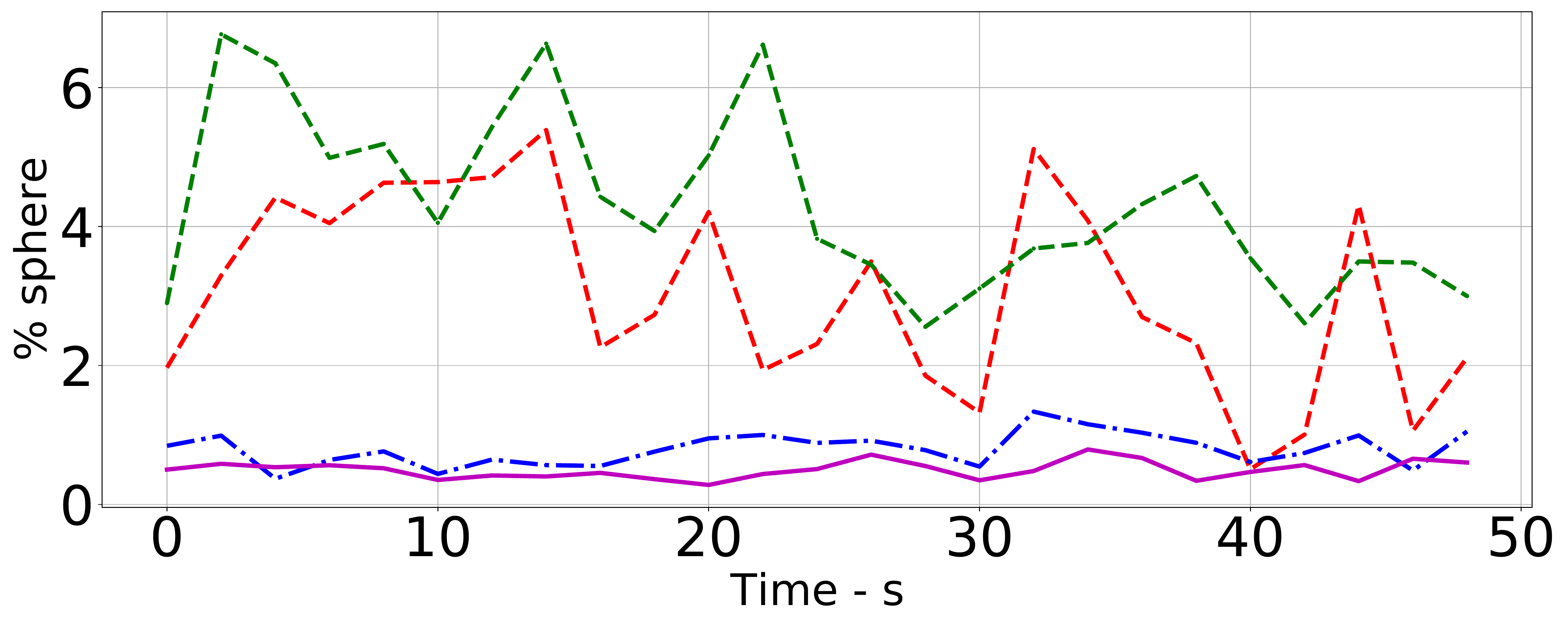}%
    \label{fig:percent sphere rollercoaster}}
\subfloat[Pairwise \% sphere diff. -timelapse ]{\includegraphics[width=0.244\textwidth]{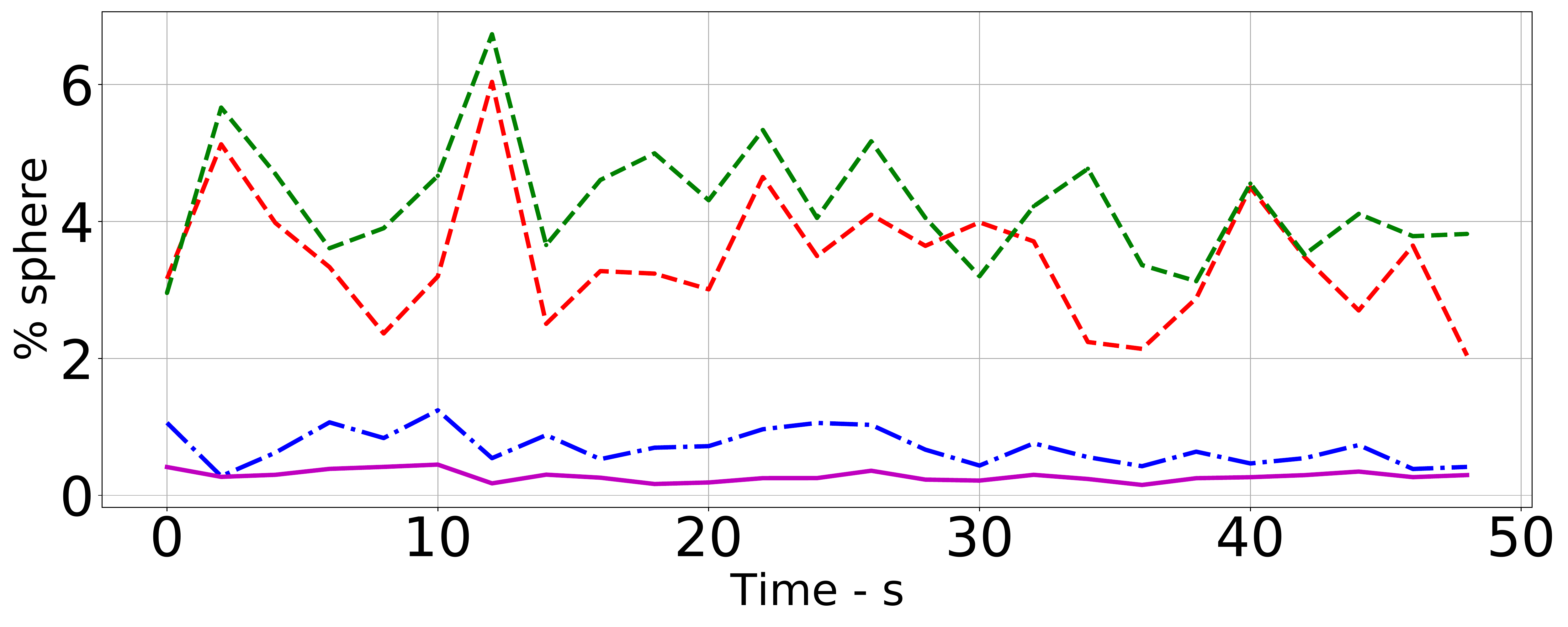}%
    \label{fig:percent sphere timelapse}}
\vspace{-4mm}
\caption{Comparison of the viewport (VP) clustering algorithms for Rollercoaster and Timelapse videos. }
\vspace{-2mm}
\label{fig:Temporal VP clustering vs proposed}
\end{figure}

Fig.~\ref{fig:Temporal VP clustering vs proposed} contains a temporal comparison of the 4 viewport clustering algorithms for two videos picked from the dataset; Rollercoaster and Timelapse. 
Majority of \textit{Roller-coaster} users fixate on the center of the frame while \textit{Time-lapse} users explore the entire frame freely without a significant correlation. The number of clusters generated by each algorithm at each time point is indicated on the corresponding plot on Fig.~\ref{fig:VP overlap rollercoaster}~and~\ref{fig:VP overlap timelapse}. Since Spherical clustering\cite{rossi_spherical_2019} shows the best performance out of the three algorithms used for comparison, we used the same number of clusters yielded by Spherical clustering for our algorithm at each time bin. 

Spherical clustering algorithm produces significantly high number of clusters as it is based on maximal clique clustering. 
Nevertheless, for the same number of clusters, the proposed algorithm yields almost on par performance in terms of viewport overlap.
Furthermore, our proposed algorithm significantly outperforms all others in pairwise speed and pairwise \% sphere explored metrics by explicitly using those features to define the viewprot trace chunk. The literature suggests that having viewport traces with similar speeds, orientation and spatial distribution within a cluster facilitates more accurate and long term viewport prediction. \textit{Viewport prediction followed by clustering is not in the scope of this paper.}

\textit{\textbf{Takeaway-} Unlike other algorithms that rely primarily on the spatial details of the viewports, the proposed feature extraction and clustering method successfully identifies clusters of viewport trace chunks that not only attend to the same content, but are also similar in behavioral aspects such as the speed of head of movement, maximum angular displacement within the window, and the extent of the sphere viewed. Moreover, the proposed algorithm is fast and lightweight in comparison to Trajectory based\cite{petrangeli_trajectory-based_2018} and Spherical clustering\cite{rossi_spherical_2019} approaches that utilize algorithms with a high computational overhead.}

\vspace{-2mm}
\subsection{Viewport clustering for aggregated dataset}\label{subsection:Viewport clustering for the entire dataset}

As the first step of our dynamic video categorization, we generated clusters inside the set of all viewport trace chunks $\{t_{i,k,j}\}$ as introduced in Table\ref{tab:basic notation}. The same feature vectors derived in \ref{Section 4- feature extraction}, $\underline{f}^{( 1)}_{i,k,j}$ were used as inputs to the K-Means clustering algorithm. 

\begin{figure}[h]
\centering
\subfloat[.]{\includegraphics[width=0.235\textwidth]{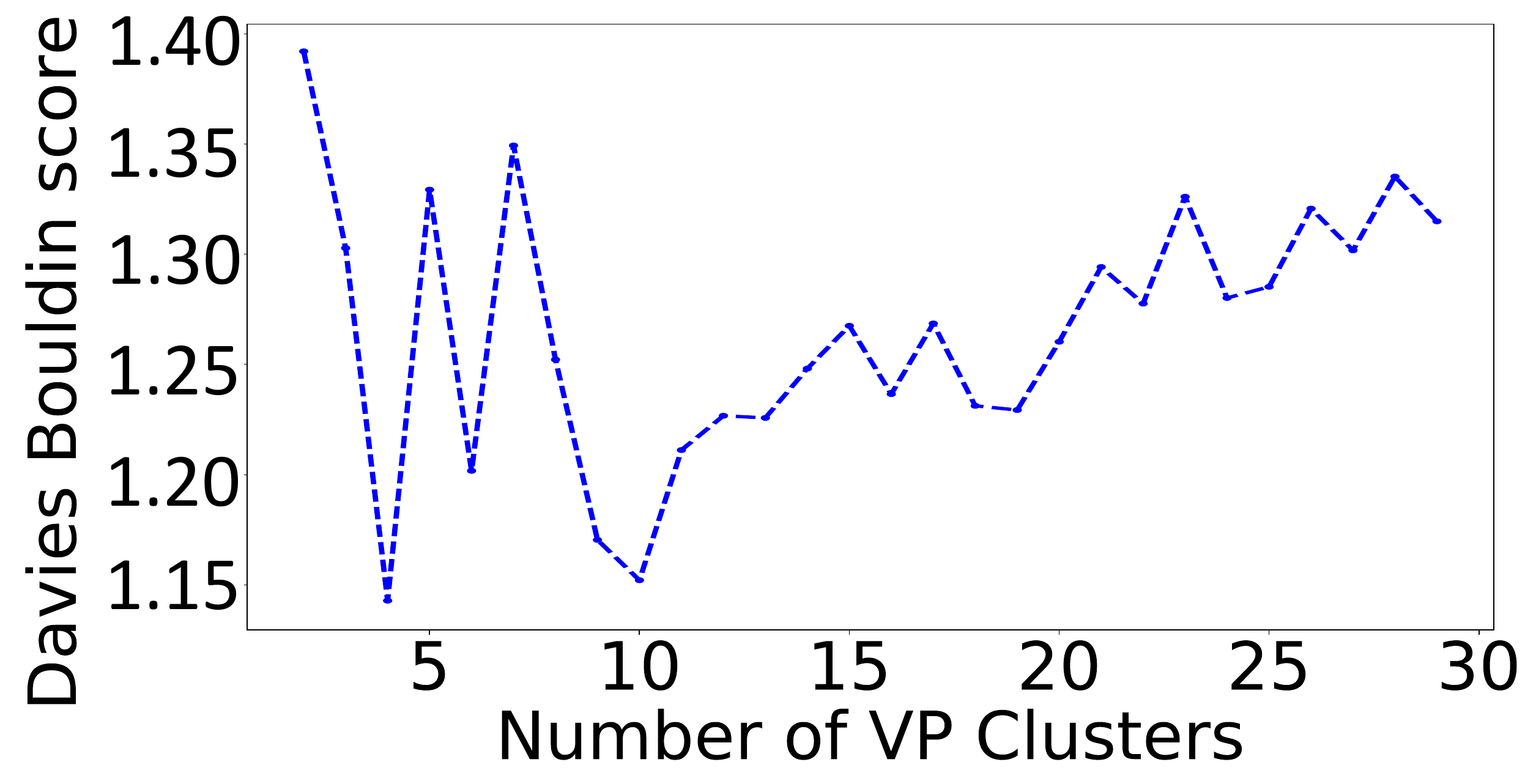}
    \label{fig:DBScore analysis VP}}
\subfloat[.]{\includegraphics[width=0.245\textwidth]{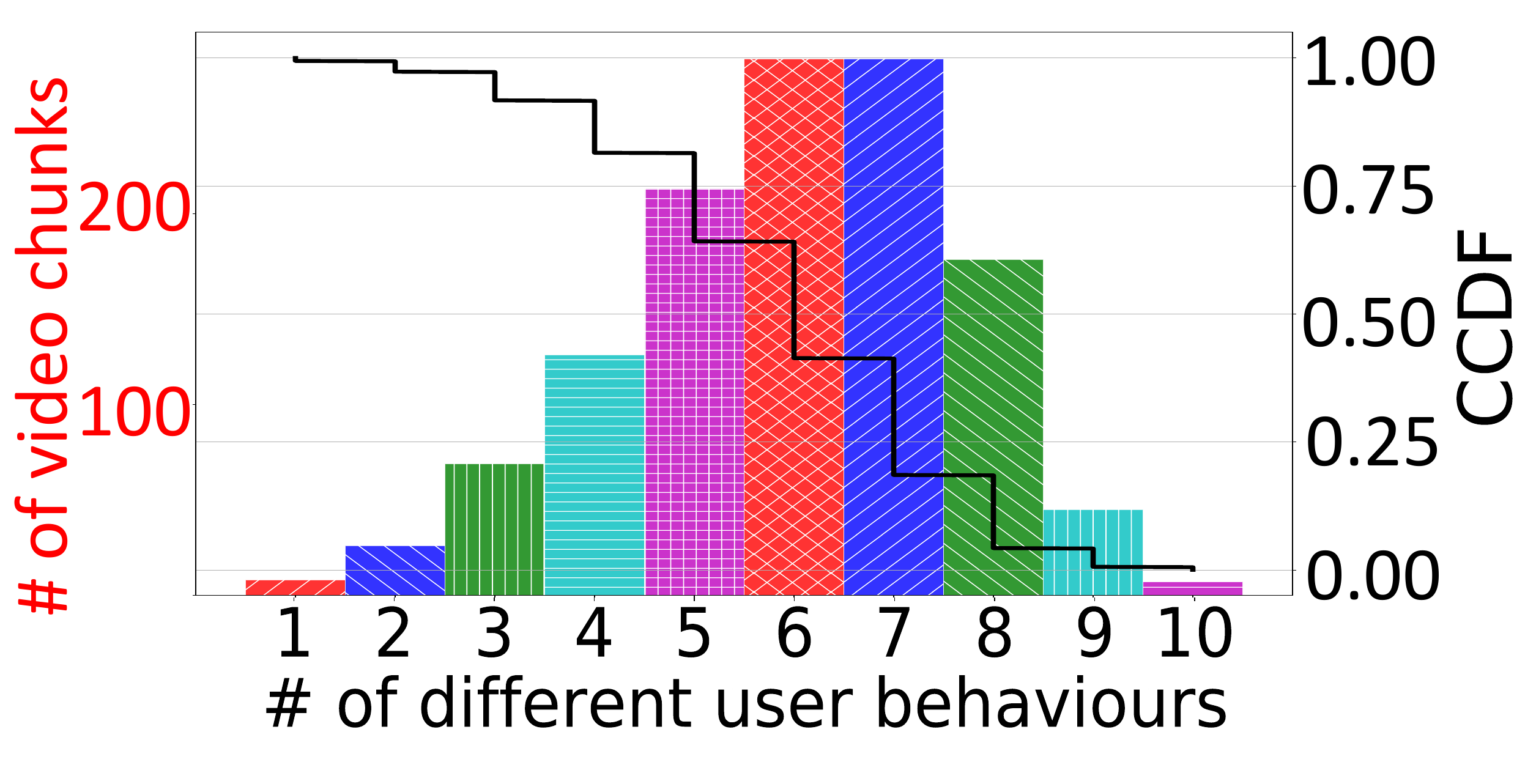}
    \label{fig:Behaviours per video chunk}}\\
\vspace{-4mm}
\caption{a)-DB-score analysis to determine the optimum number of clusters for viewport trace chunk clustering. b)-Number of different user behaviors available in video chunks and the CDF of their distribution}\vspace{-5mm}
\label{fig:Similarity measurement VP m=10}
\end{figure}

\begin{figure}[h]
\centering
\subfloat[M=3 ]{\includegraphics[width=0.16\textwidth, , angle=90]{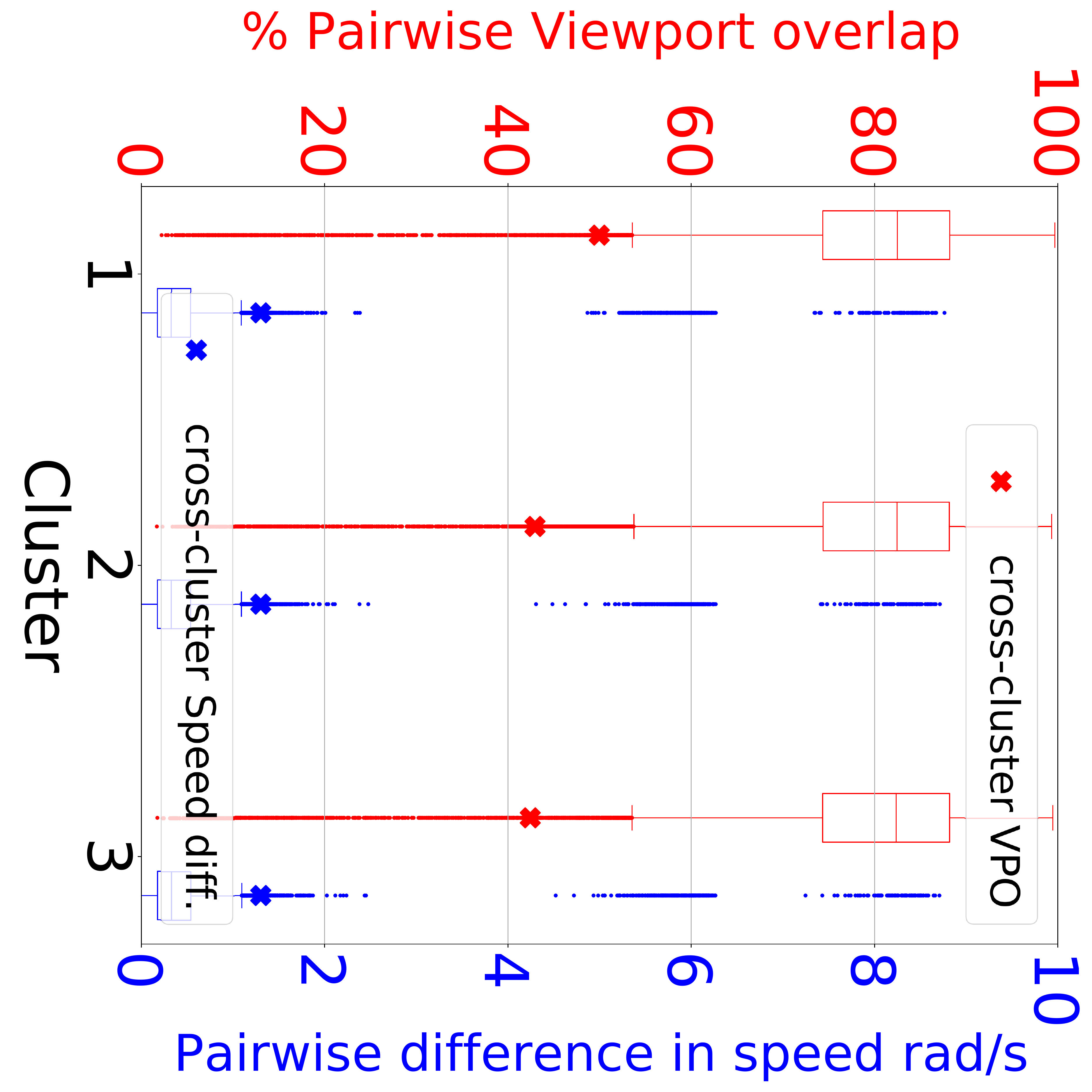}
    \label{fig:Similarity measurement within clusters for M=3}}
\subfloat[M=10 ]{\includegraphics[width=0.34\columnwidth, , angle=90]{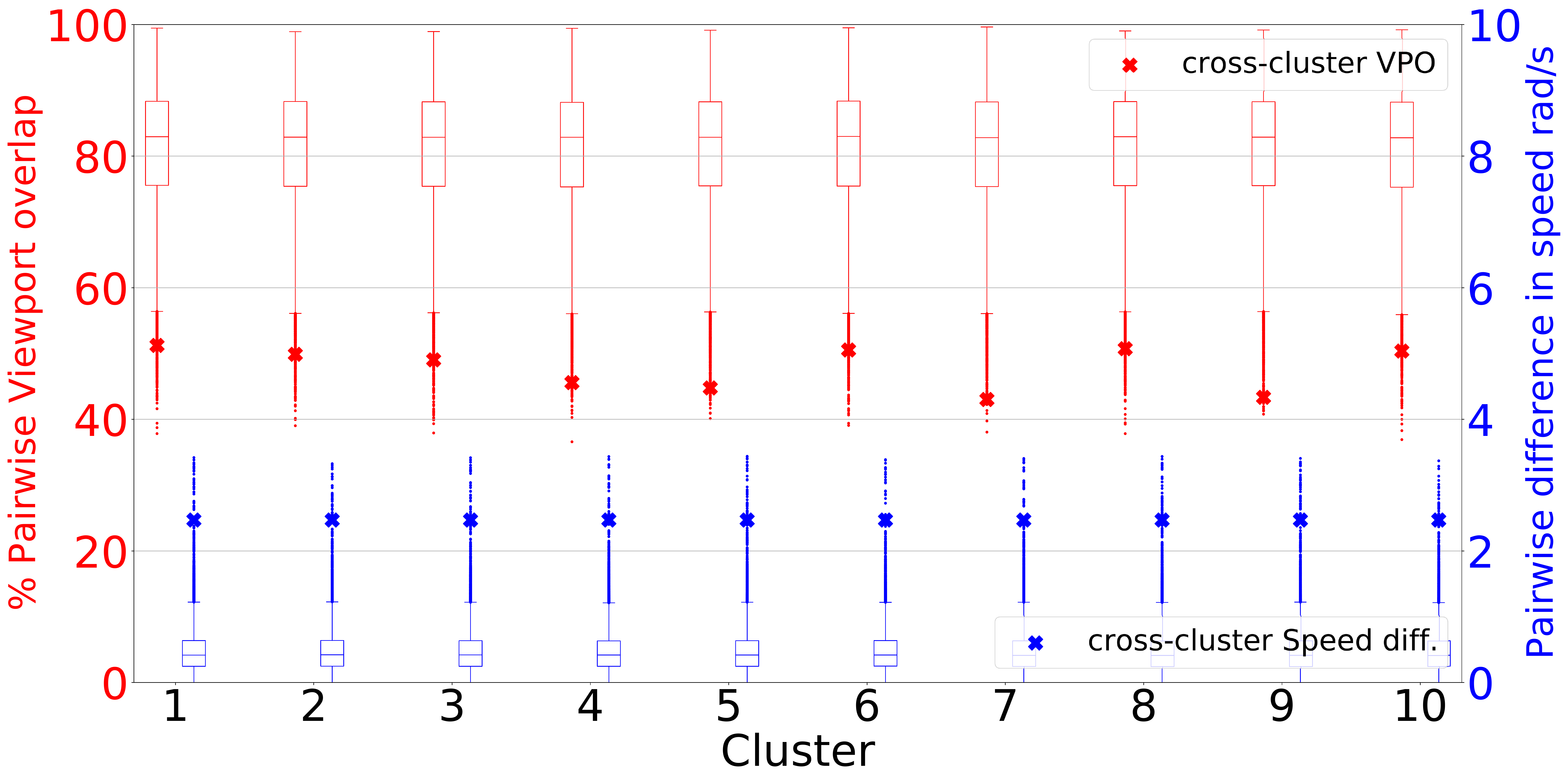}
    \label{fig:Similarity measurement within clusters for M=10}}\\
\vspace{-4mm}
\caption{Pairwise VPO and Pairwise difference in head movement speed within and between clusters for \textit{M=3, 10}}
\vspace{-4mm}
\label{fig:Similarity measurement VP m=10}
\end{figure}

The optimum number of clusters for the set $\{t_{i,k,j}\}$ was determined by a DB (Davies- Bouldin) Score analysis as indicated in Fig.~\ref{fig:DBScore analysis VP}. The graph gives 2 minimums; \textit{M}=3 and \textit{M}=10. According to the plot, \textit{M}=3 is not a stable minimum and as shown in Fig.~\ref{fig:Similarity measurement within clusters for M=3}~and~\ref{fig:Similarity measurement within clusters for M=10}, upon clustering the dataset to 3 clusters, 2.29\% of the datapoints were identified as outliers. Where as for 10 clusters the outlier percentage was only 1.12\%. \textbf{Therefore, we selected \textit{M}=10}.

\vspace{-2mm}

Fig.~\ref{fig:Similarity measurement VP m=10} illustrates the results for similarity measurement within and between viewport clusters generated by the proposed algorithm. 
We observe that for each cluster, the resulting percentage pairwise viewport overlap between the traces of a cluster is significantly higher than that between traces from different clusters. 
The averaged pairwise viewport overlap within viewport clusters is 81.17\%, and that between different clusters is 47.80\%.
Moreover, the pairwise difference of viewport trace speeds within a cluster is significantly less than that between different clusters. While the averaged pairwise speed difference within clusters is 0.47 rad/s, that between different clusters is as high as 2.47 rad/s 
It is evident that the proposed viewport clustering algorithm captures similarities in terms of both spatial and motion features.

\vspace{-2mm}
\begin{figure}[h!]
\centering
\subfloat[Viewport trace cluster 7]{\includegraphics[width=0.22\textwidth]{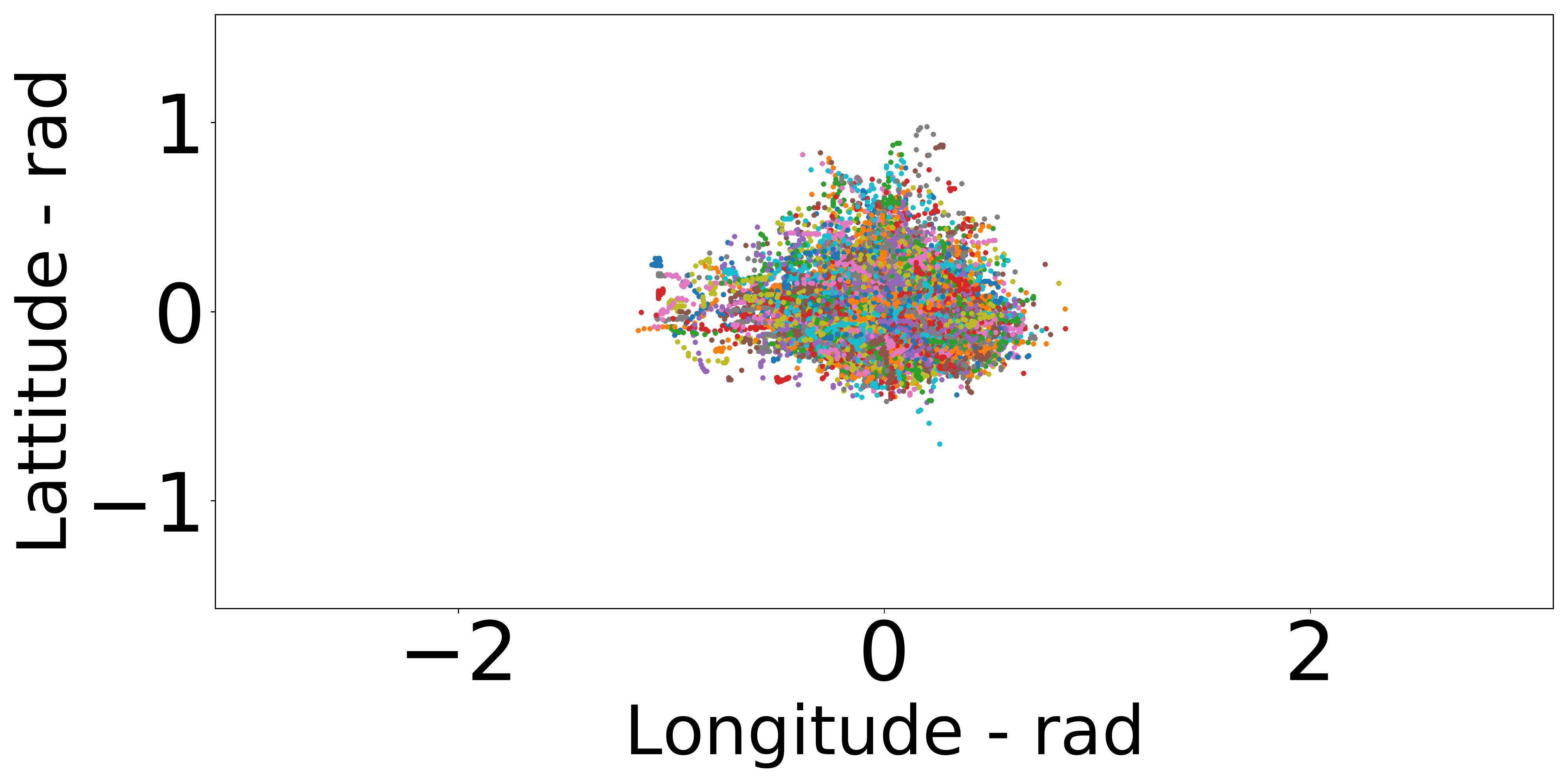}
    \label{fig:Viewport trace cluster 7}}
\subfloat[Viewport trace cluster 9]{\includegraphics[width=0.22\textwidth]{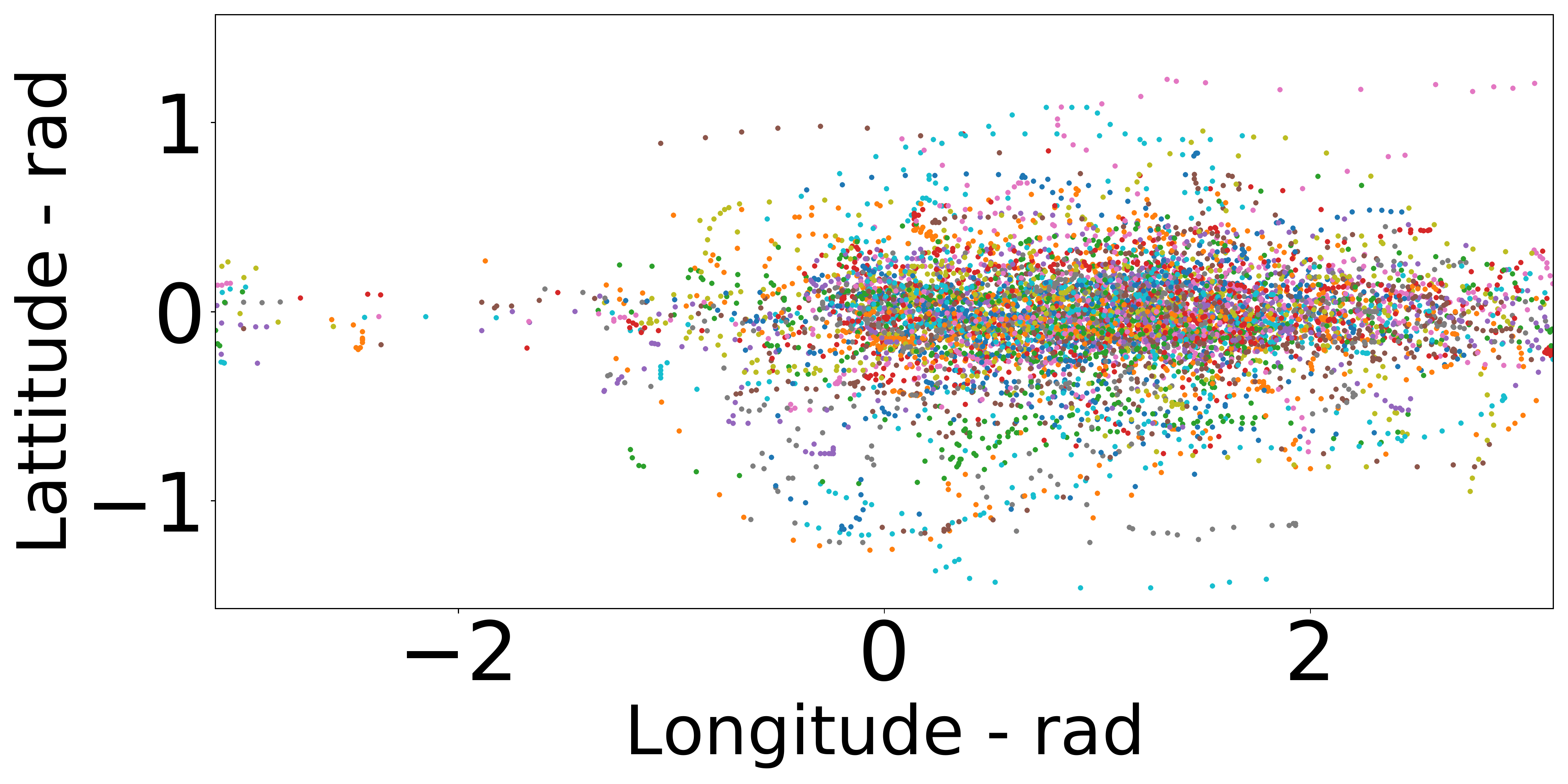}
    \label{fig:Viewport trace cluster 9}}
    \hfill

\subfloat[Cluster 7 features]{\includegraphics[width=0.18\textwidth]{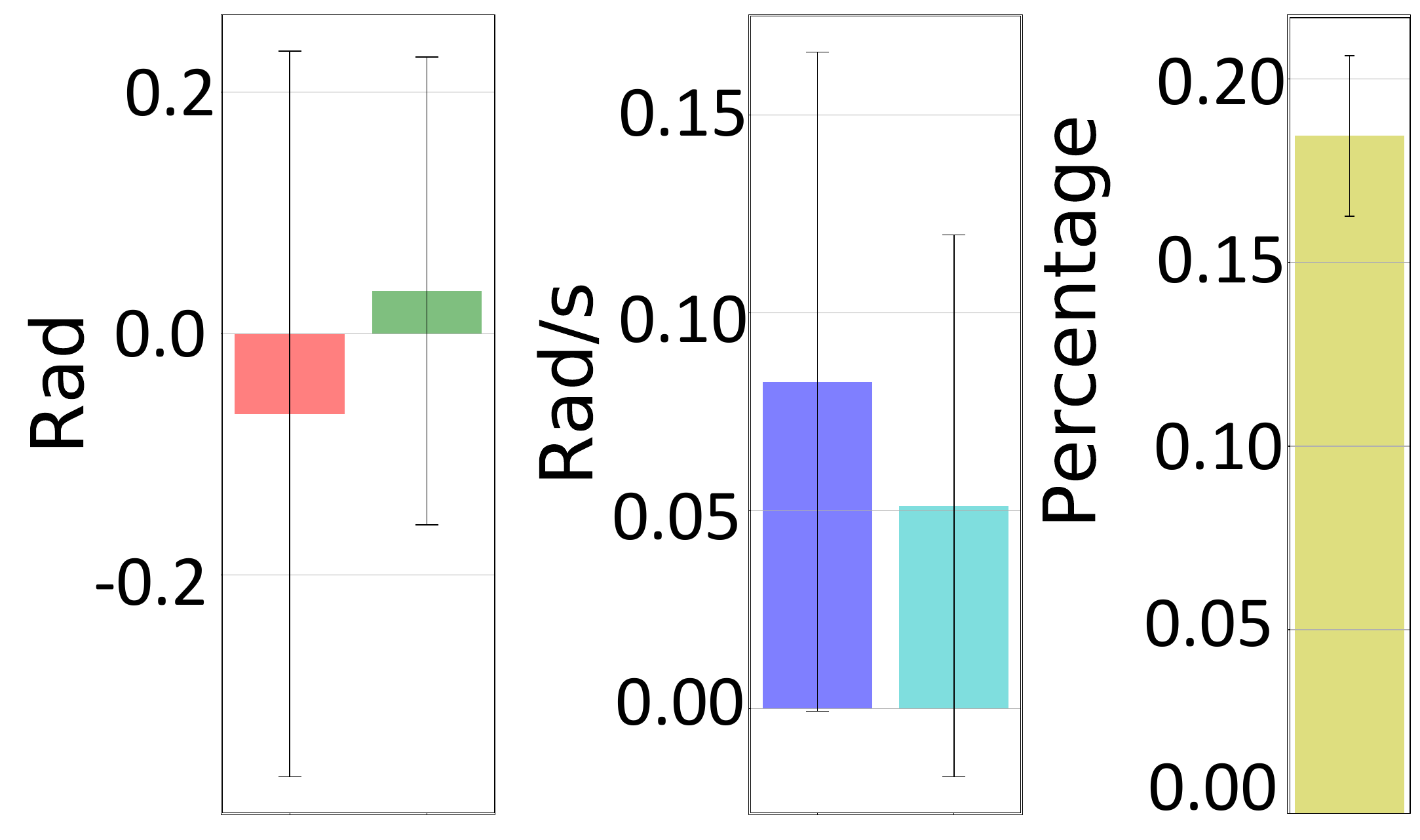}
    \label{fig:features cluster 7}}
\subfloat[Cluster 9 features]{\includegraphics[width=0.26\textwidth]{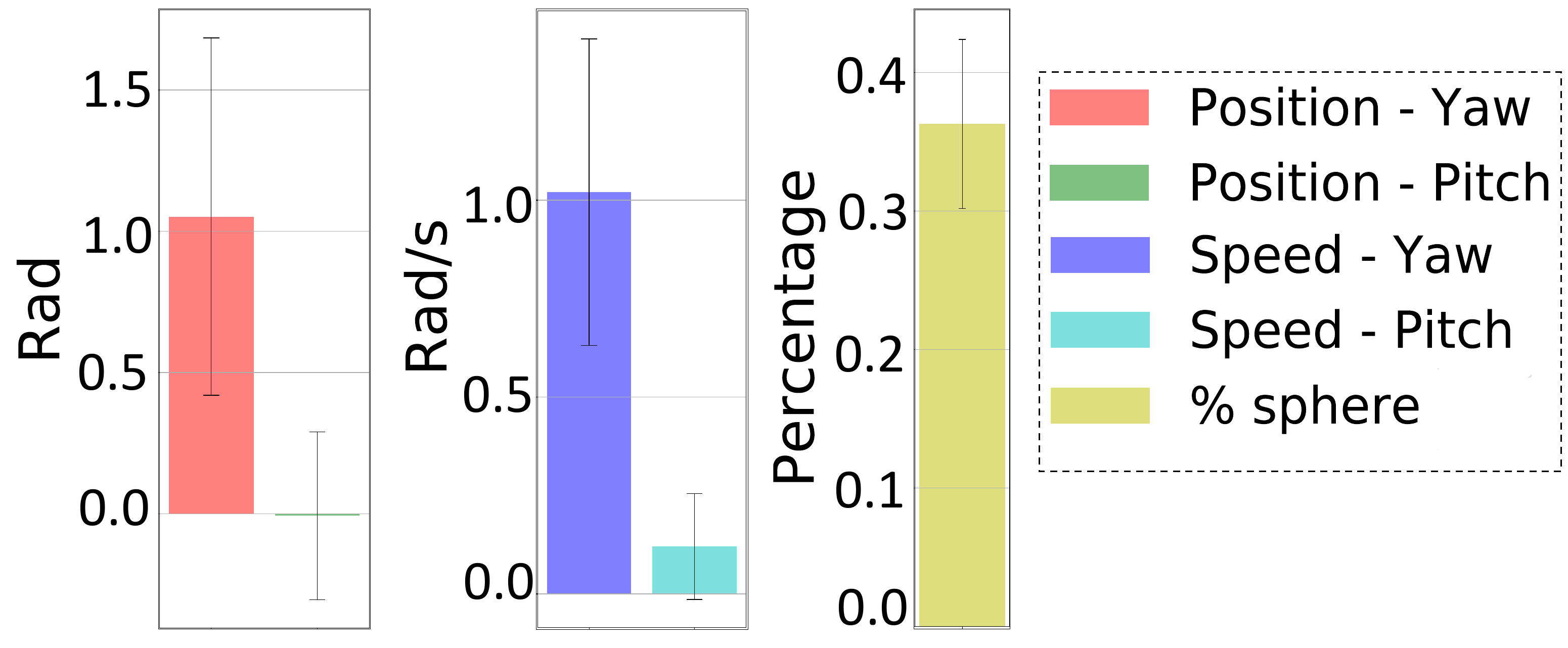}
    \label{fig:features cluster 9}}
    \hfill
\vspace{-3mm}
\caption{Viewport distributions and corresponding features of clusters 7 and 9}
\vspace{-3mm}

\label{fig:Viewport distributions of different clusters}
\end{figure}

Fig.~\ref{fig:Viewport trace cluster 7} and \ref{fig:Viewport trace cluster 9} show example intermediate results of viewport clustering indicating viewport trace distribution on Equirectangular frame for clusters 7 and 9. Each colored dot indicates one sample of viewport centers of a given user in the cluster. A viewport trace for 2s is composed of 20 such points. 
It is evident how different user behaviors are captured by different clusters. Cluster 7 contains user traces that are slowly moving or are stationary around the center of the frame while cluster 9 contains users that are moving fast in the equatorial axis biased to the right of the frame.  
Each cluster can contain traces with different combinations of features describing position and motion as shown in in Fig.~\ref{fig:Viewport distributions of different clusters}.
Fig.~\ref{fig:Behaviours per video chunk} shows the number of different viewport trace clusters (different behaviors) to which the users of a particular video chunk belongs. To eliminate the effects of outliers, a behavior is counted only if there are two or more users belonging to that particular cluster. According to the CCDF, More than 75\% of all video chunks have users displaying more than 4 different user behavior types (out of 10). 

\textit{\textbf{Takeaway-} Different users consuming different \ang{360} videos show similarities in terms of positioning, speed, and \% of sphere explored by their viewports. The novel feature definition successfully identifies such similar behaviors. Majority of the video chunks have users displaying a variety of user trace patterns. Hence, for a successful characterization of videos, all different user behaviors should be used }

\begin{figure}[t!]
\centering
\includegraphics[width=0.8\columnwidth]{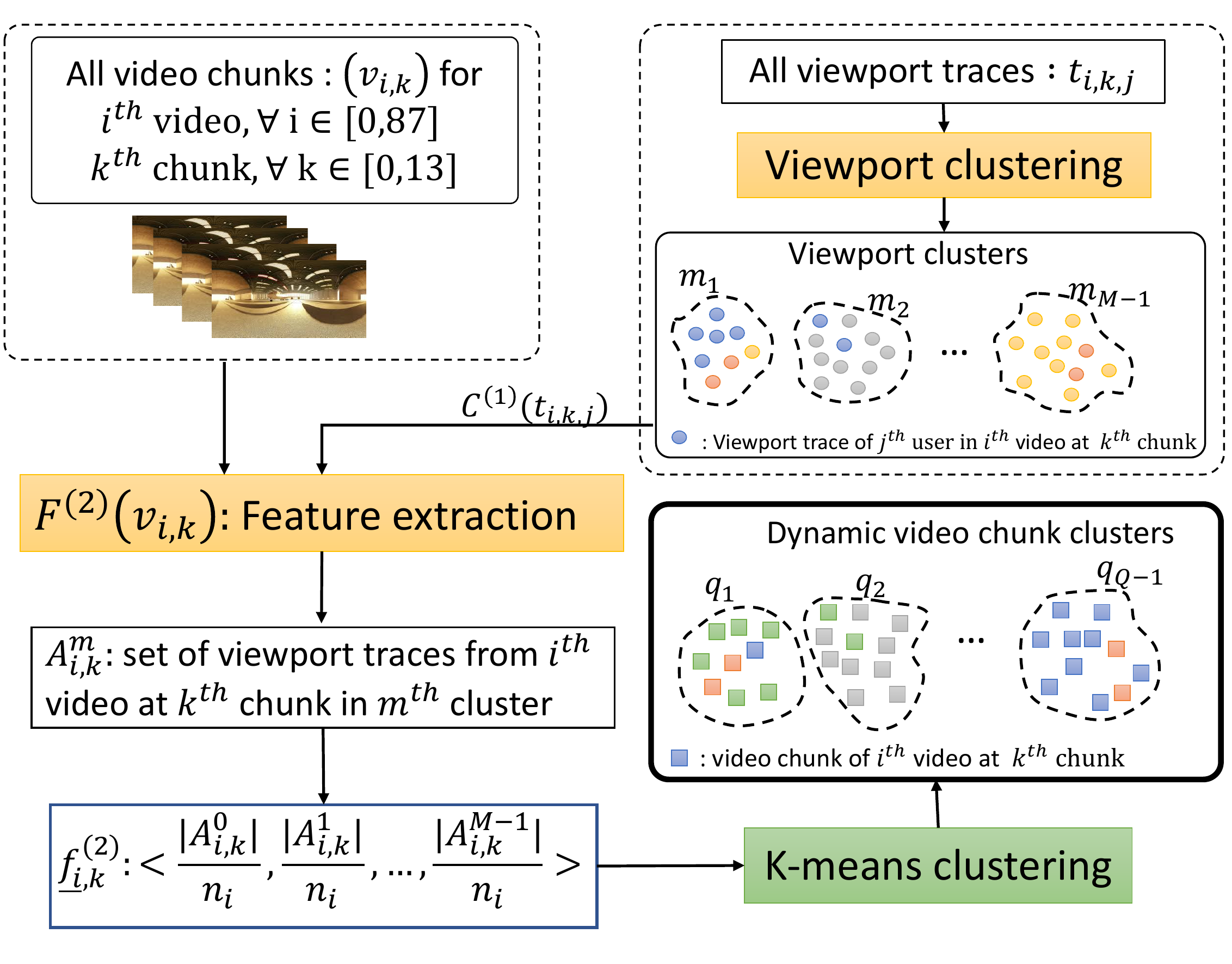}%
\label{fig:viewport_clustering_based_categorization}
\vspace{-4mm}
\caption{Overview of dynamic video categorization}
\label{fig:overview_dynamic_video_cat}\vspace{-5mm}
\end{figure}

\begin{figure*}[t!]
\centering

{\includegraphics[width=0.161\textwidth]{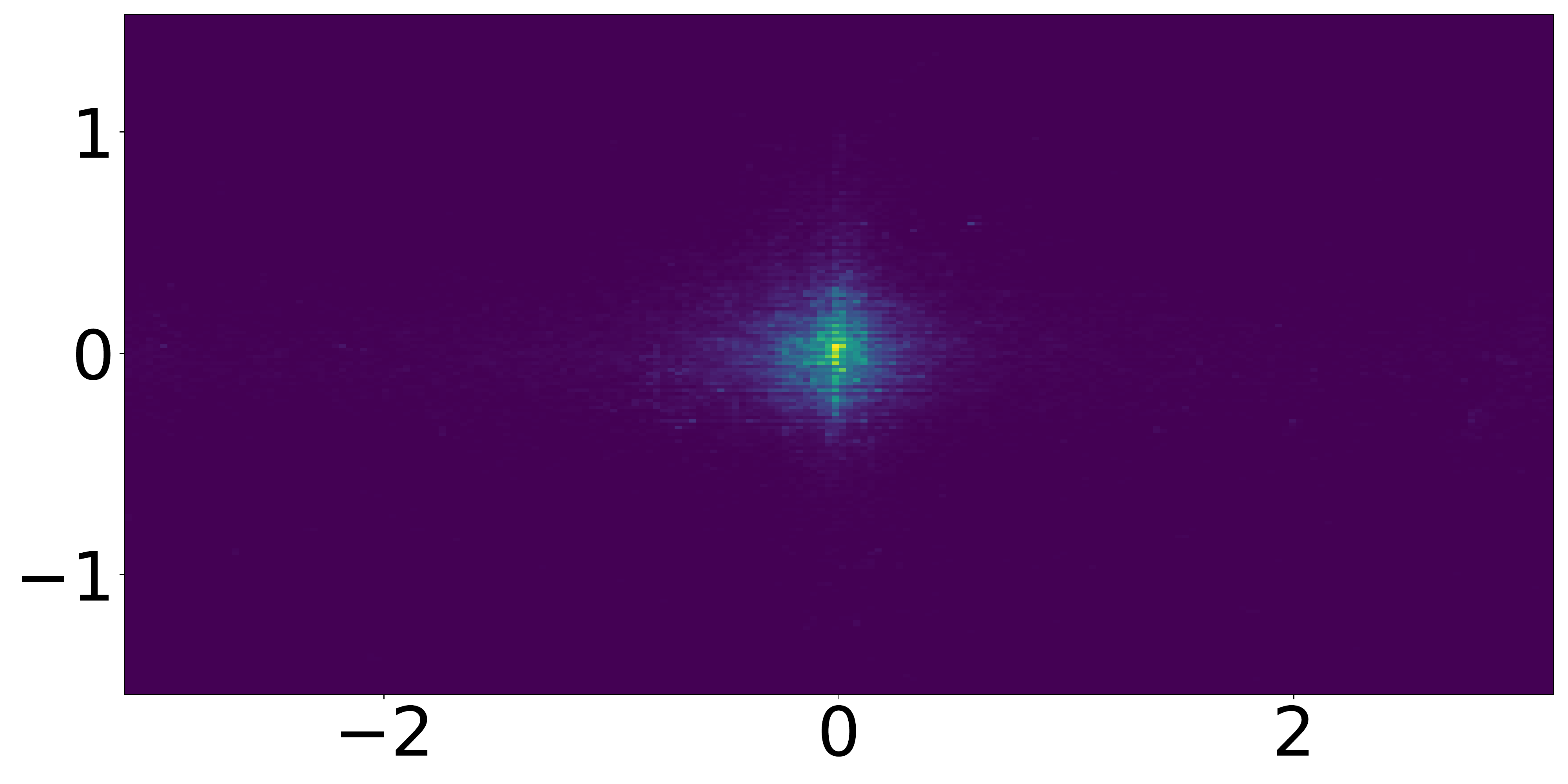}
}
{\includegraphics[width=0.161\textwidth]{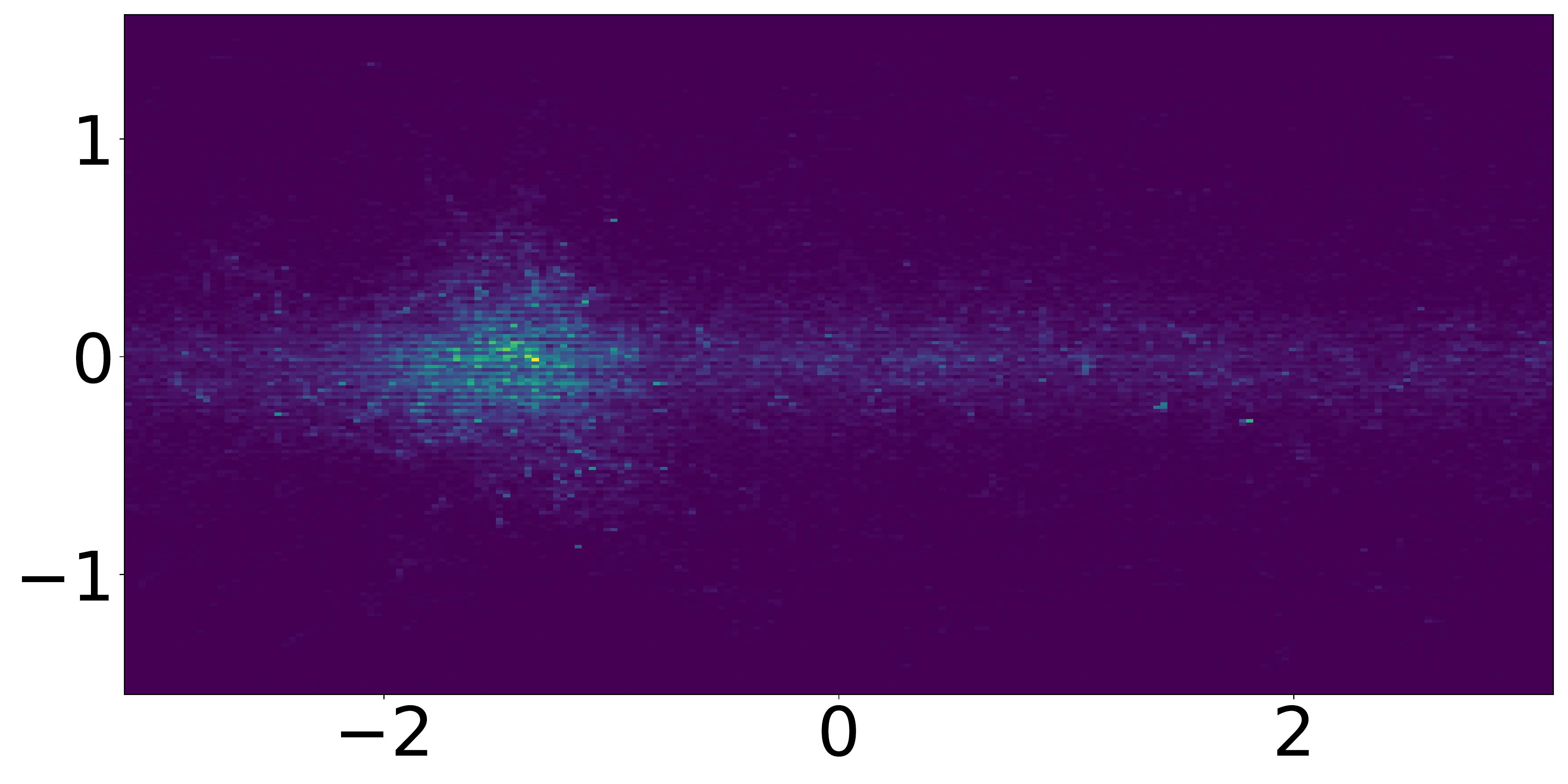}
}
{\includegraphics[width=0.161\textwidth]{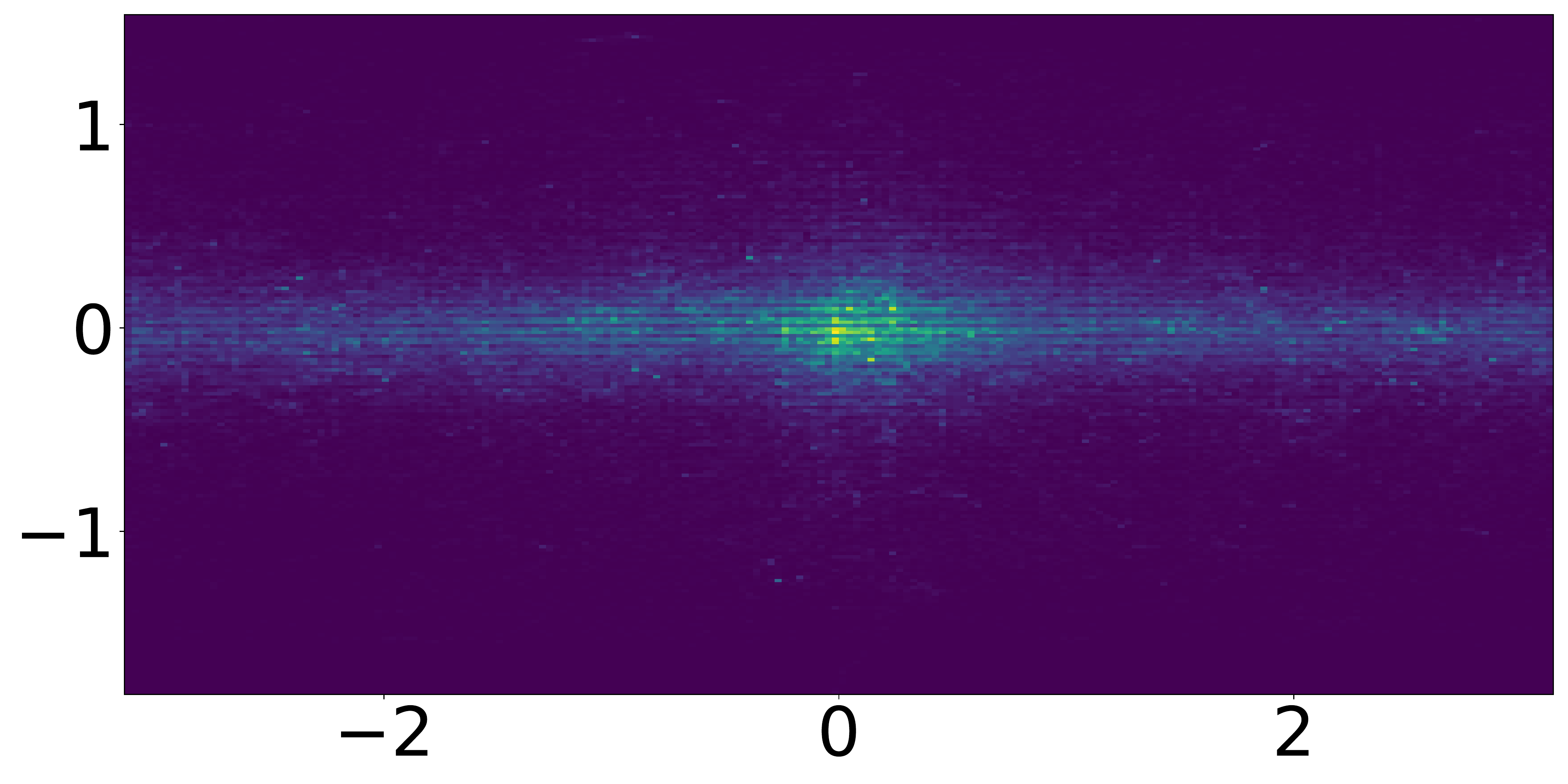}}
{\includegraphics[width=0.161\textwidth]{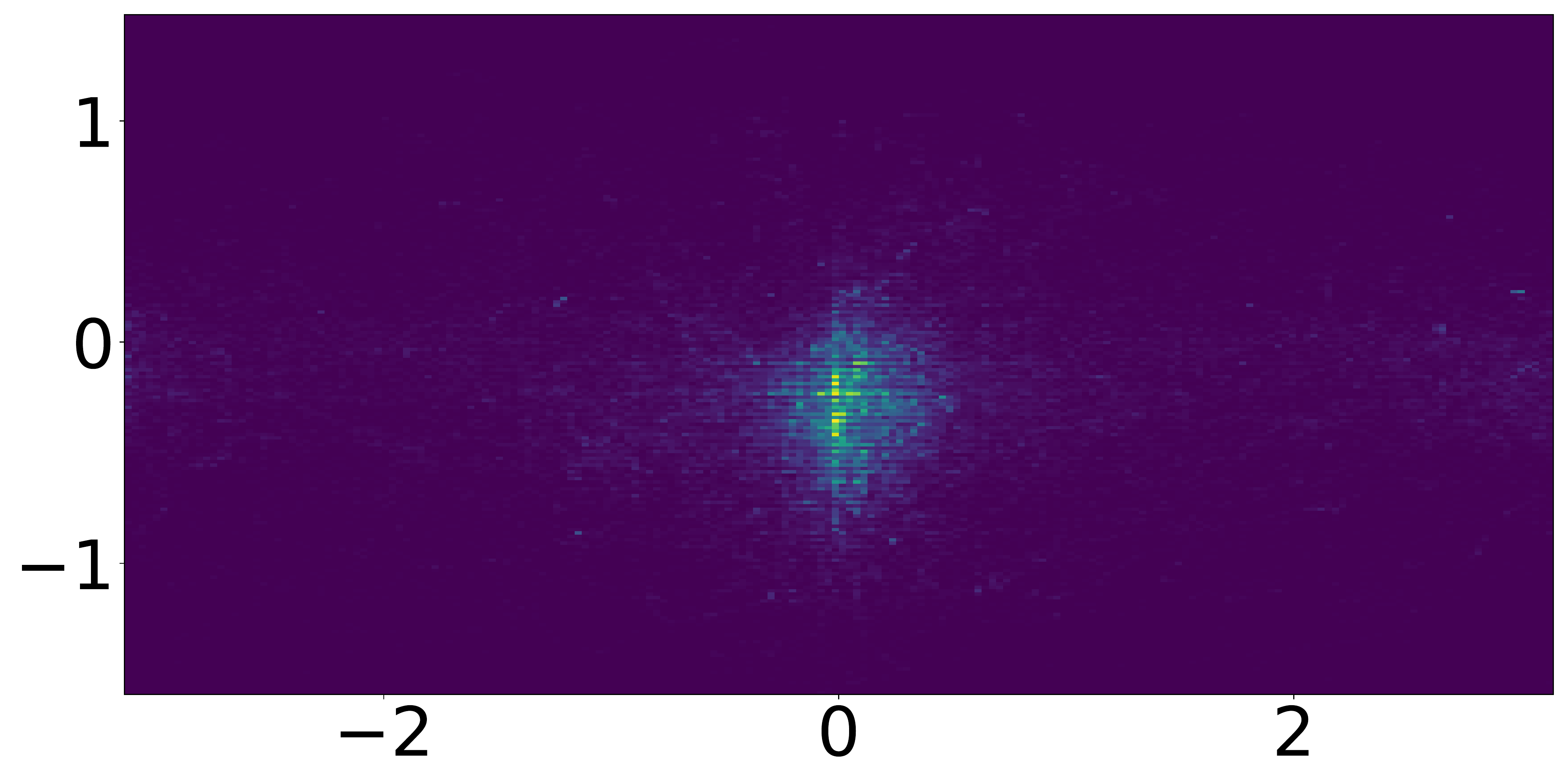} }
{\includegraphics[width=0.161\textwidth]{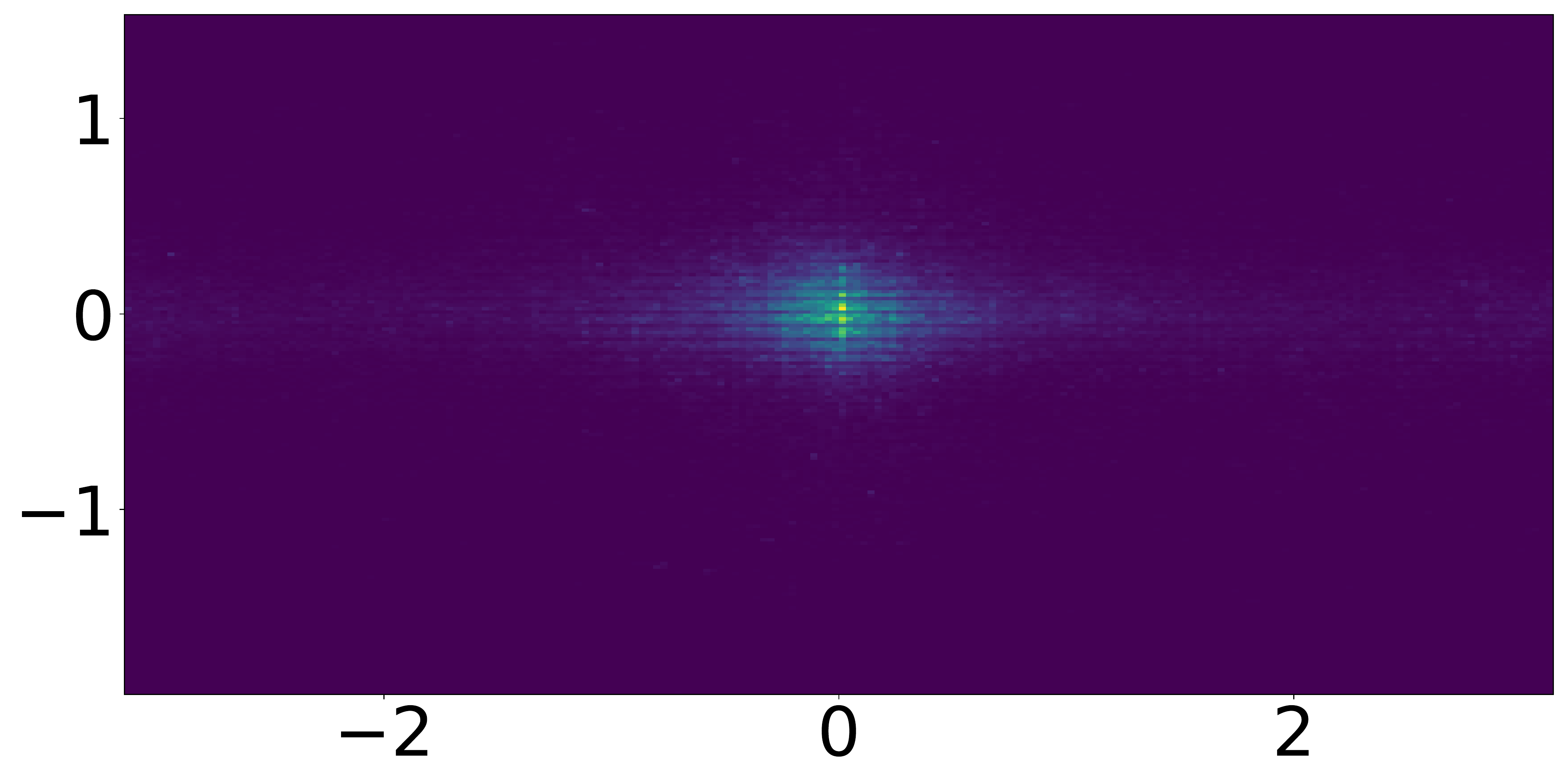}}
{\includegraphics[width=0.161\textwidth]{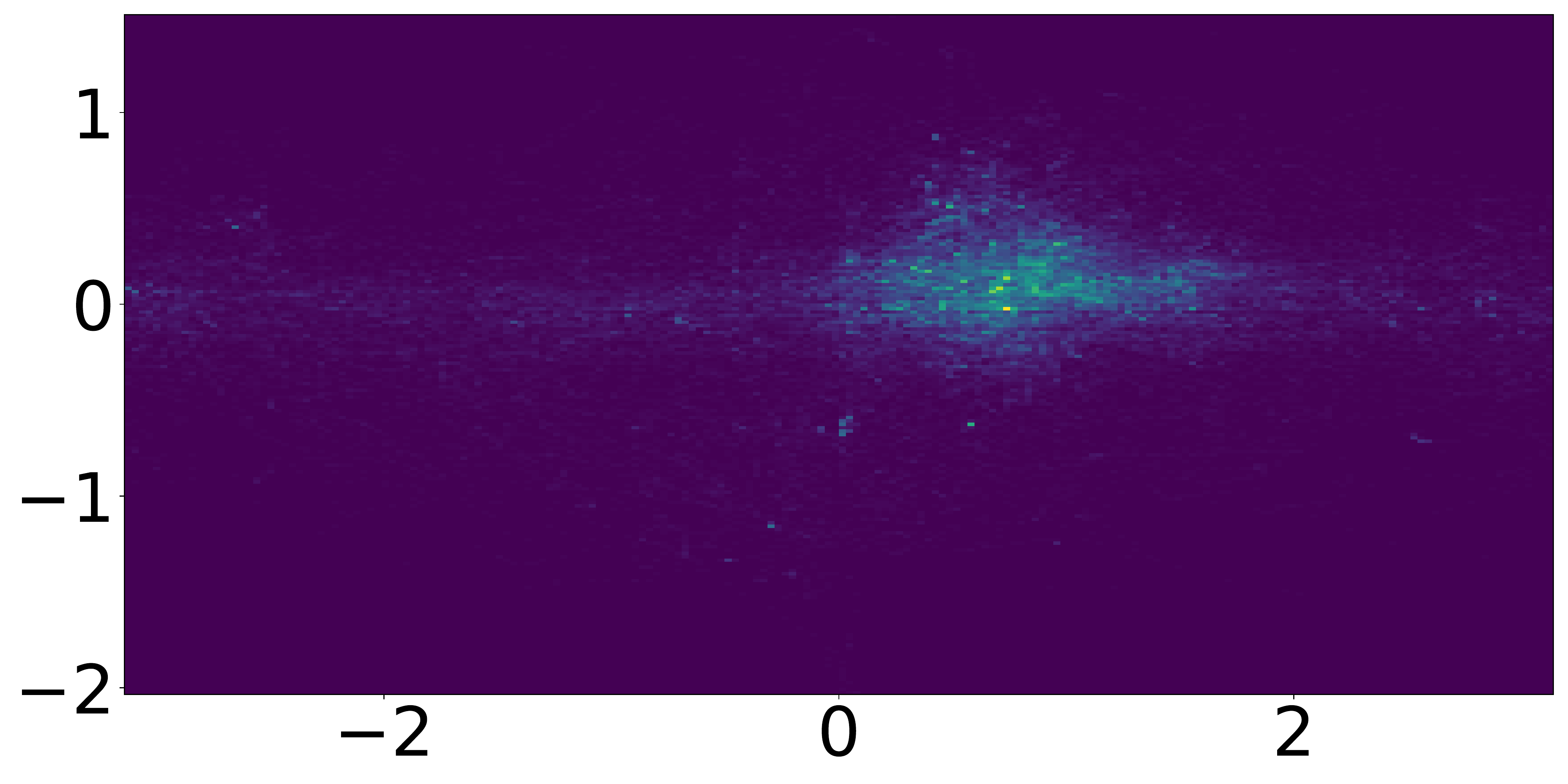}}

{\includegraphics[width=0.161\textwidth]{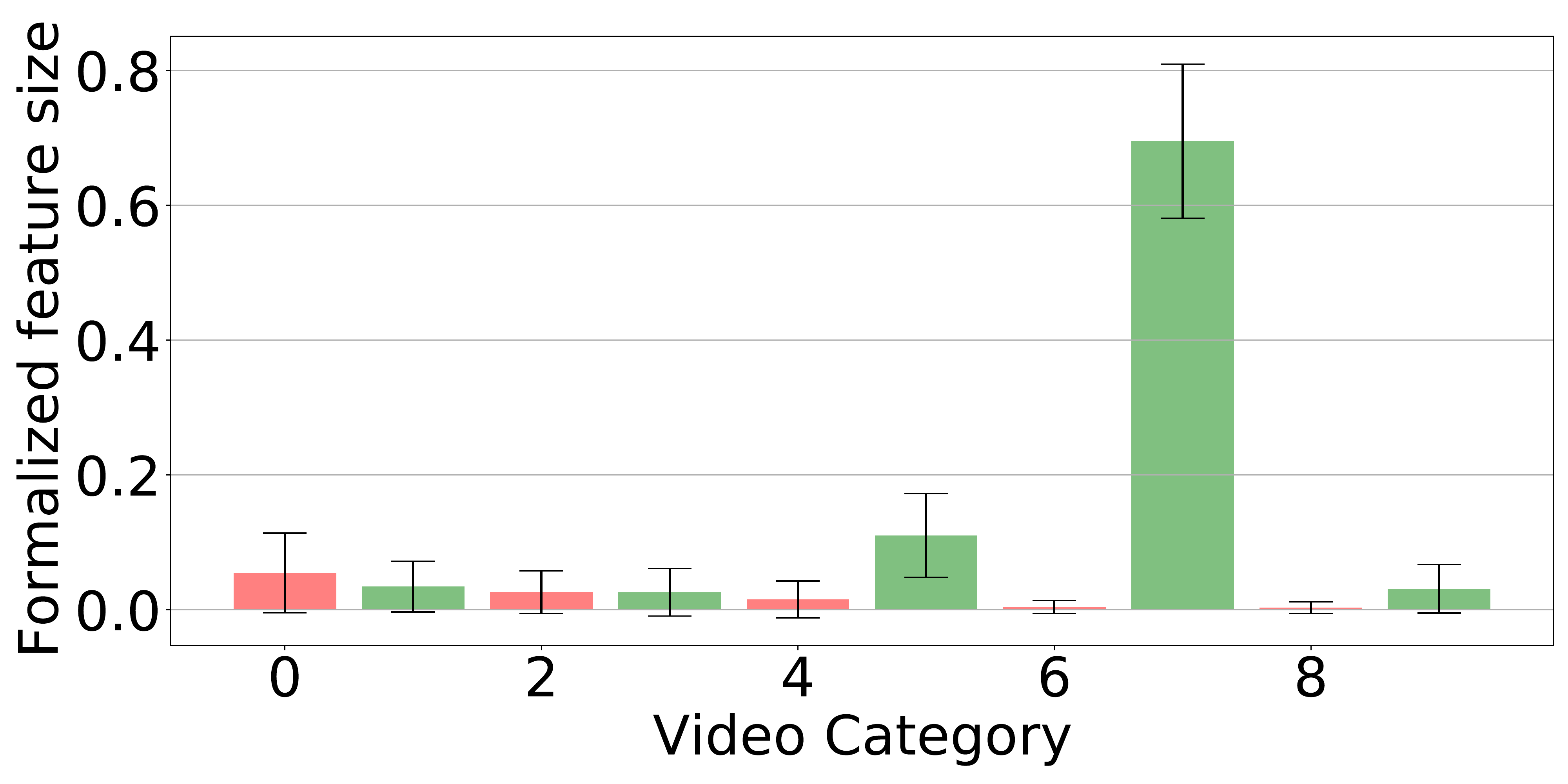}
}
{\includegraphics[width=0.161\textwidth]{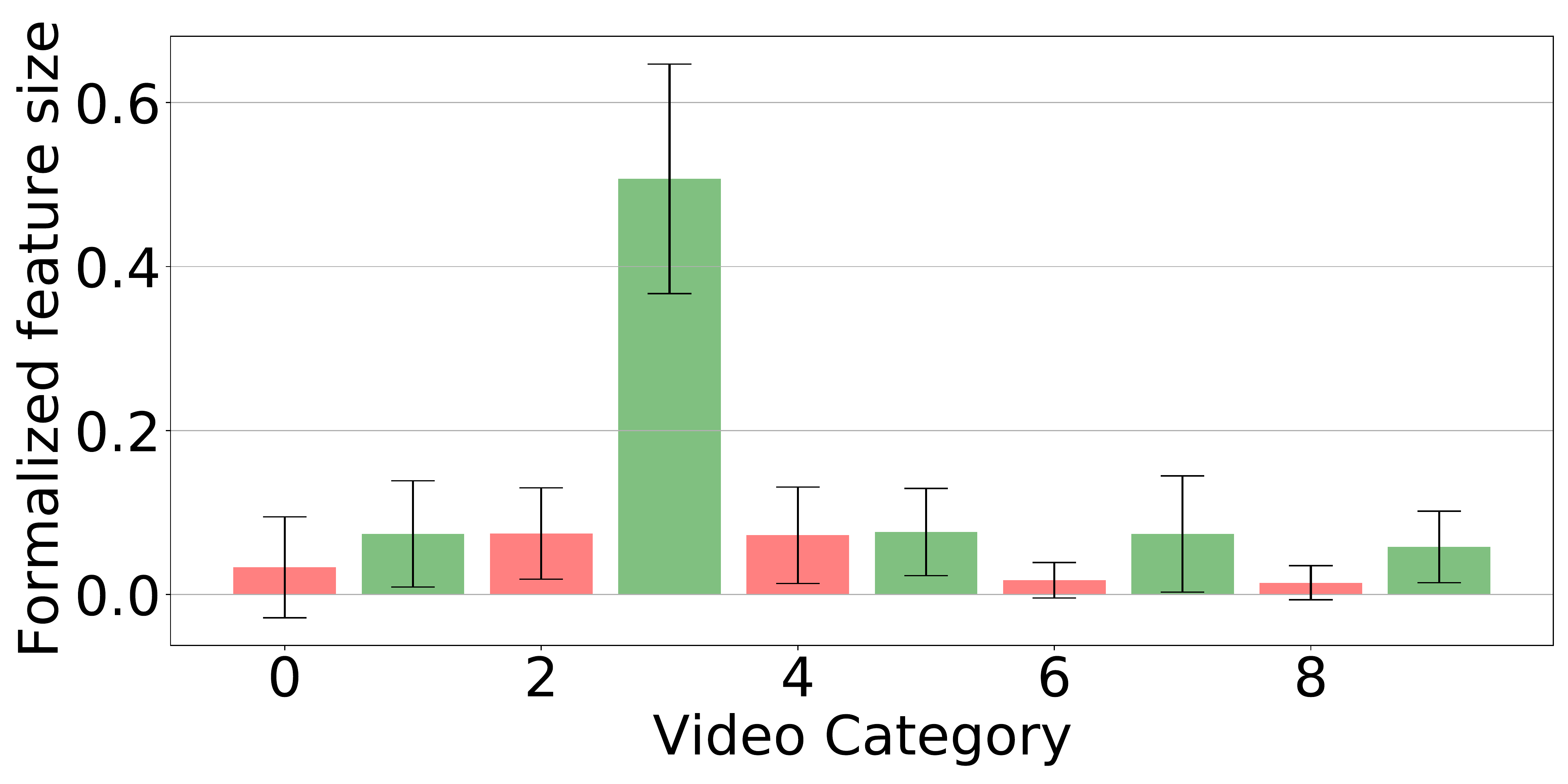}
}
{\includegraphics[width=0.161\textwidth]{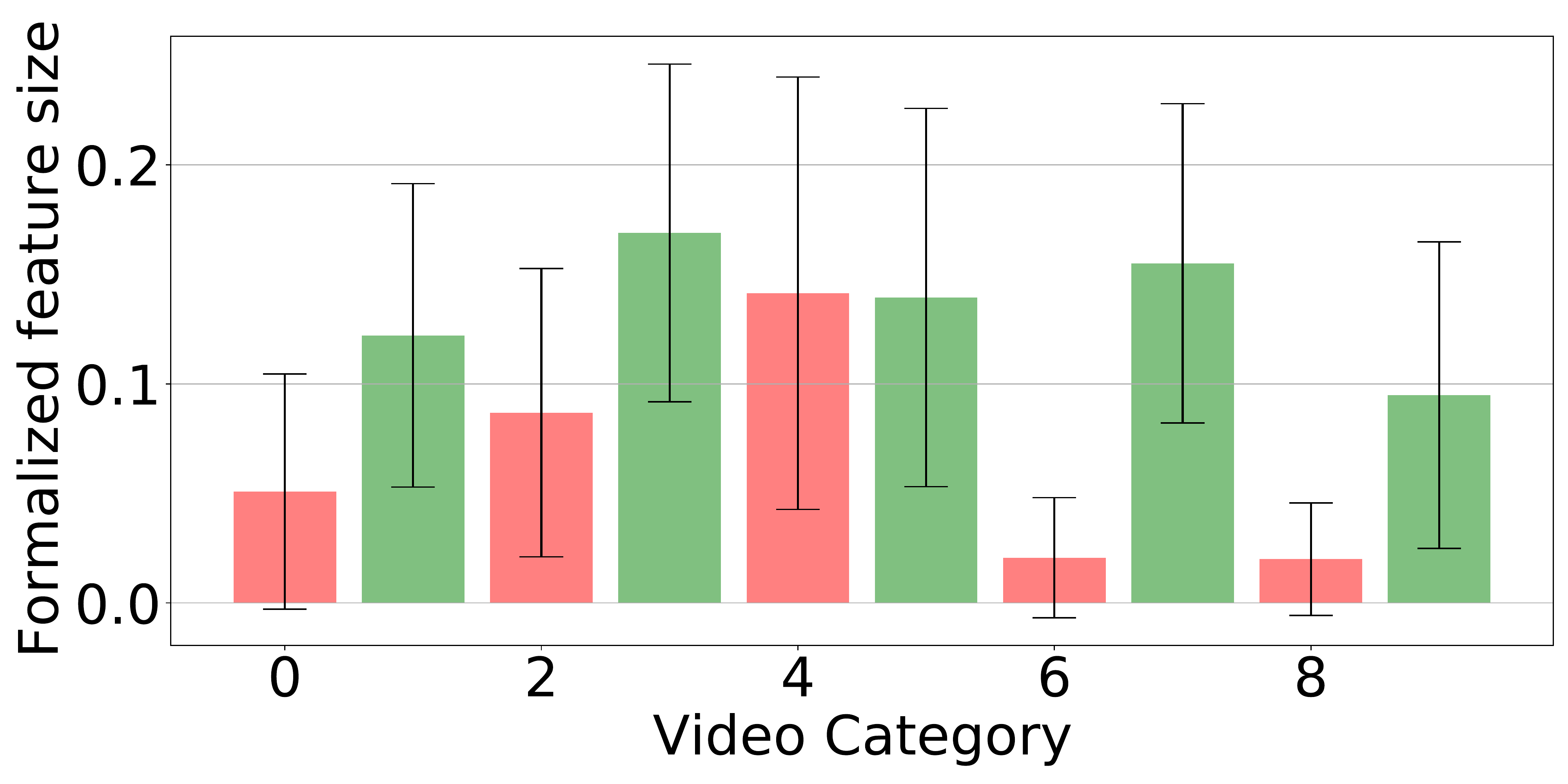}}
{\includegraphics[width=0.161\textwidth]{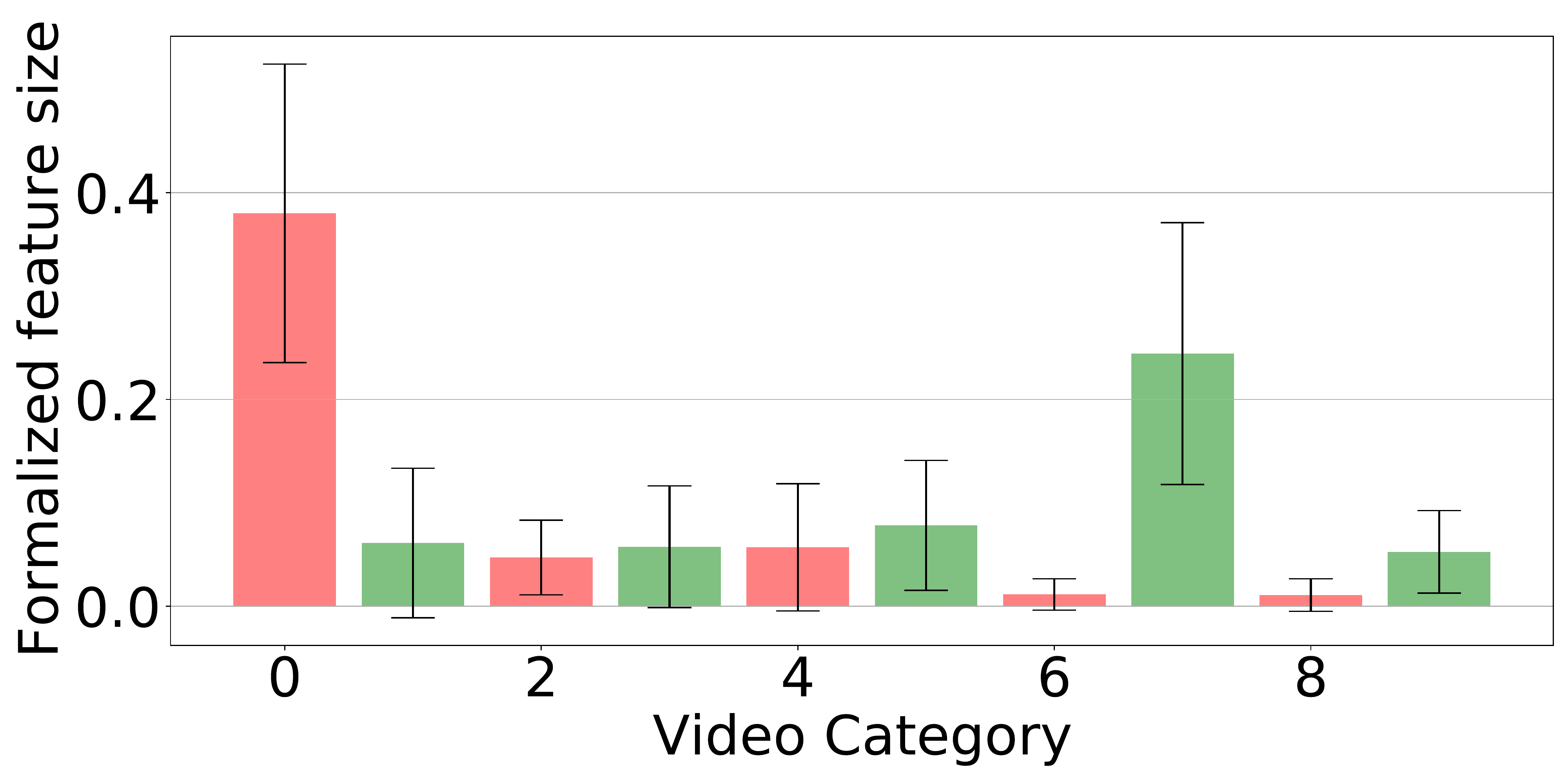} }
{\includegraphics[width=0.161\textwidth]{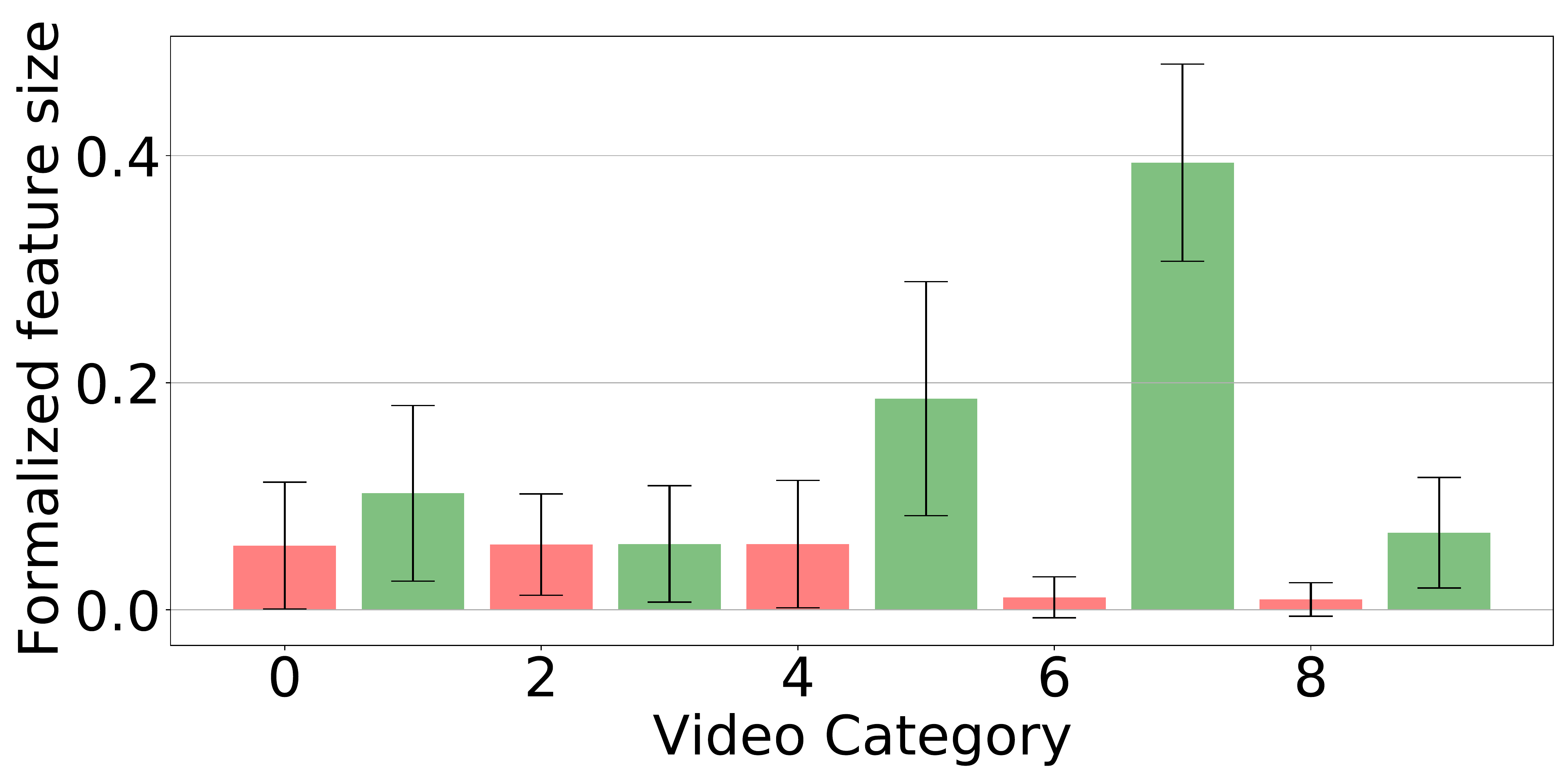}}
{\includegraphics[width=0.161\textwidth]{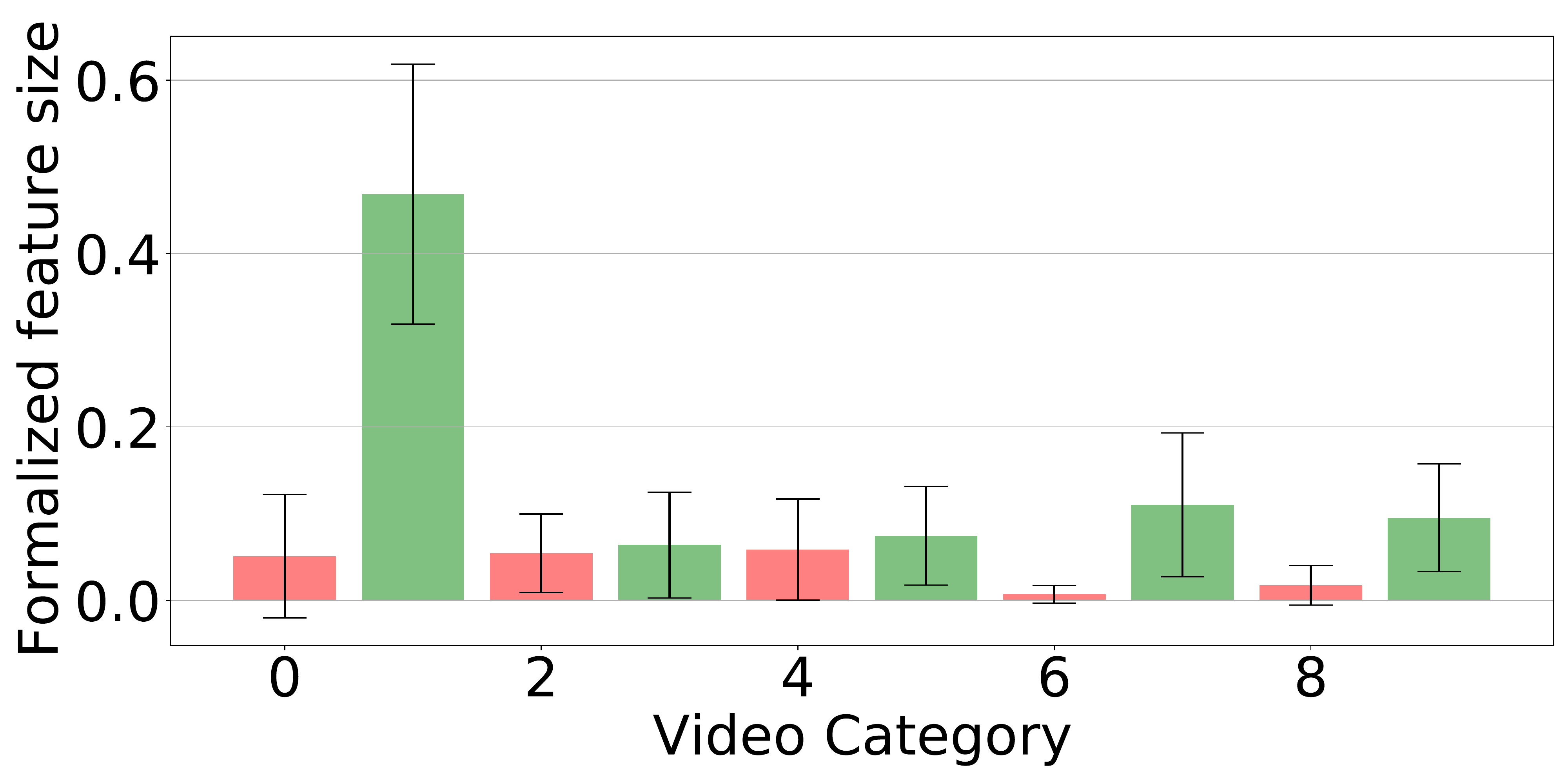}}
\vspace{-6mm}
\caption{The heatmaps represent the distribution of viewport trace chunks in each of the resulting video categories. The histograms represent the average distribution of features as given in Eq.~\ref{eq:vp_traces_vector_4_vid_clustering} for each category}\vspace{-5mm}
\label{fig:Video categories heatmaps and feature hist}
\end{figure*}

\vspace{-2mm}
\section{Dynamic Video chunk categorization} \label{section 5}

In this section, we present a novel algorithm to dynamically categorize video chunks. Inspired by the results of Section \ref{subsection:Viewport clustering for the entire dataset}, we statistically characterise a video chunk by the distribution of how the users perceive its content by using different viewport trace patterns. We start off with the hypothesis that \textit{for a large enough dataset, the aggregate set of all user viewport trace chunks can reasonably represent all practical types of short-term user behaviors in perceiving 360°videos.} 
In the following subsections we define a feature representation for video chunks based on our results from section \ref{subsection:Viewport clustering for the entire dataset}, categorize the video chunks into optimum number of clusters, and evaluate the performance of the proposed categorization. 

\vspace{-2mm}
\subsection{Feature engineering for video chunks}
Let ${F^{( 2)}( v_{i,k}) =\underline{f}^{( 2)}_{i,k}}$ be the feature extraction function for video chunks. .
For a particular video chunk $v_{i,k}$ it generates an \textit{M} dimensional feature vector $\underline{f}^{( 2)}_{i,k}$ as follows. (\textit{M} is the number of viewport trace clusters as explained in section \ref{subsection:Viewport clustering for the entire dataset}. We selected \textit{M}=10.)




\vspace{-2mm}
\begin{equation}\label{eq:vp_traces_vector_4_vid_clustering}
{\underline{f}^{( 2)}_{i,k}} =<\frac{|A^{0}_{i,k} |}{n_{i}},\frac{\ |A^{1}_{i,k} |}{n_{i}},\ ...,\frac{\ |A^{M-1}_{i,k} |}{n_{i}}>
\end{equation}
\vspace{-2mm}

such that $A^{m}_{i,k} =\left\{t_{i,k,j} \ |\ C^{( 1)}( t_{i,k,j}) =m,\ \forall j\in [ 0,\ n_{i})\right\}$


In a gist, the feature representation for a particular video chunk indicates the population distribution of its users that belongs to each of the  different viewport trace clusters (representing different user trace behaviors) resulting in section \ref{subsection:Viewport clustering for the entire dataset}. For this analysis we used 1232 unique \ang{360} video chunks. (88 videos, 14 time bins)


\vspace{-2mm}
\subsection{Video chunk categorization}\label{Video chunk categorization}

Next we generate clusters of similar video chunks in the defined \textit{M} dimensional feature space as shown in Fig.~\ref{fig:overview_dynamic_video_cat}. $\underline{f}^{( 2)}_{i,k} \ \forall i\in [ 0,87] ,\ \forall k\in [ 0,14)$ are taken as inputs to a K-Means clustering algorithm that yields \textit{Q} clusters. Similar to Eq.~\ref{eq:cluster_alloction_func}, corresponding cluster/category for a video chunk is given by $C^{( 2)}( v_{i,k}) =q,\ q\in [ 0,Q)$.    


We define a category as a collection of video chunks that are similarly perceived by the users. i.e, videos with similar distributions of different user trace behaviors. The following distance metric for 2 video chunks $v_{i,k}$ and $v_{i',k'}$ quantifies the above definition.







\vspace{-2mm}
\begin{equation}\label{eq:similarity_metric2}
    S( v_{i,k} ,v_{i',k'}) =\sqrt{\sum\limits ^{M}_{m=0}\left( |A^{m}_{i,k} |-|A^{m}_{i',k'} |\right)^{2}}
\end{equation} 
\vspace{-2mm}

It calculates the euclidean distance between the feature vectors of two video chunks, quantifying the difference in how the users of two videos are distributed among the 10 identified user trace patterns/behaviors. 

\begin{figure}[h!]
\centering
\subfloat[DB-Score analysis for optimum \textit{Q} ]{\includegraphics[width=0.24\textwidth]{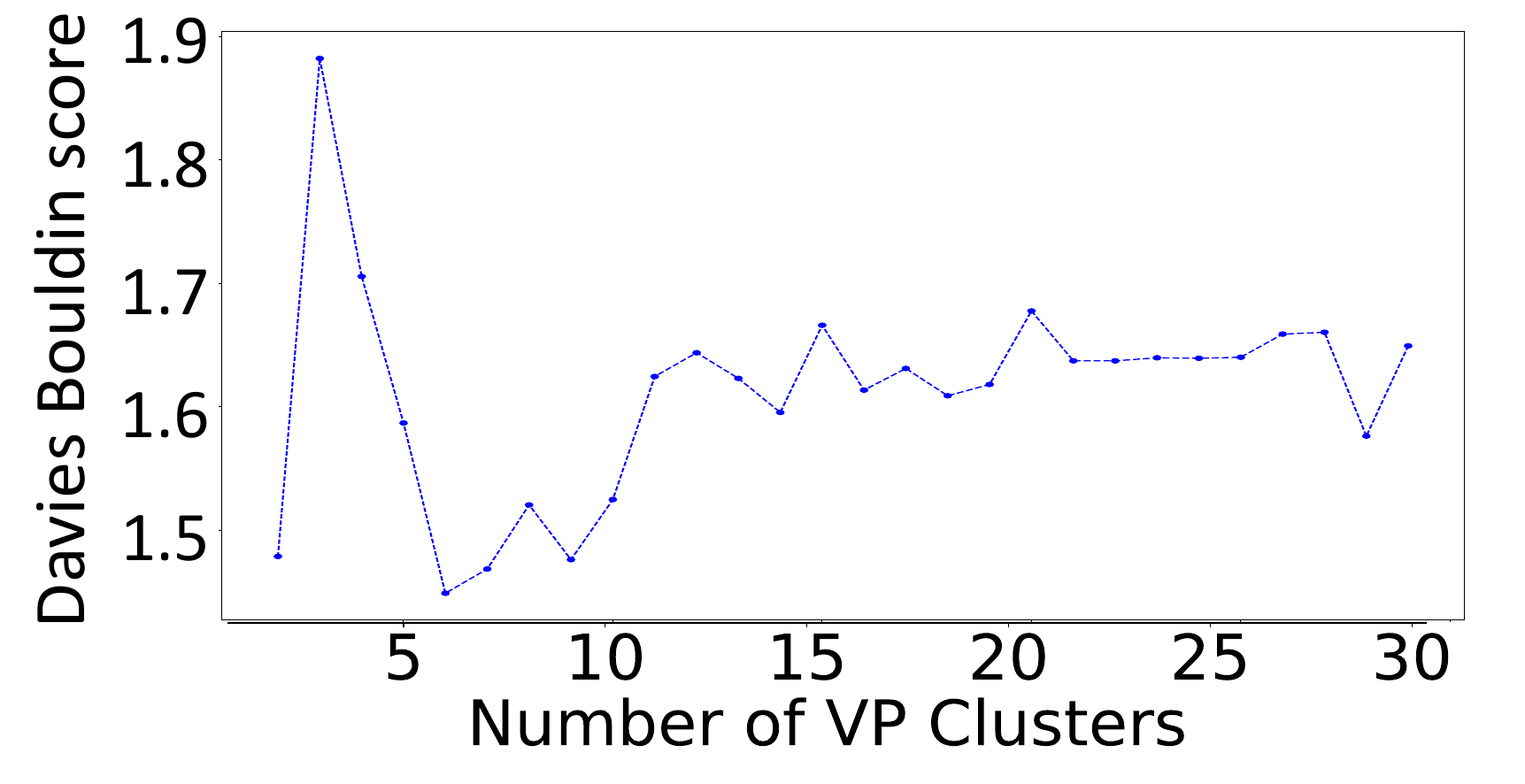}
    \label{fig:DB-Score_analysis_for_optimum_Q}}
\subfloat[Pairwise similarity]{\includegraphics[width=0.24\textwidth]{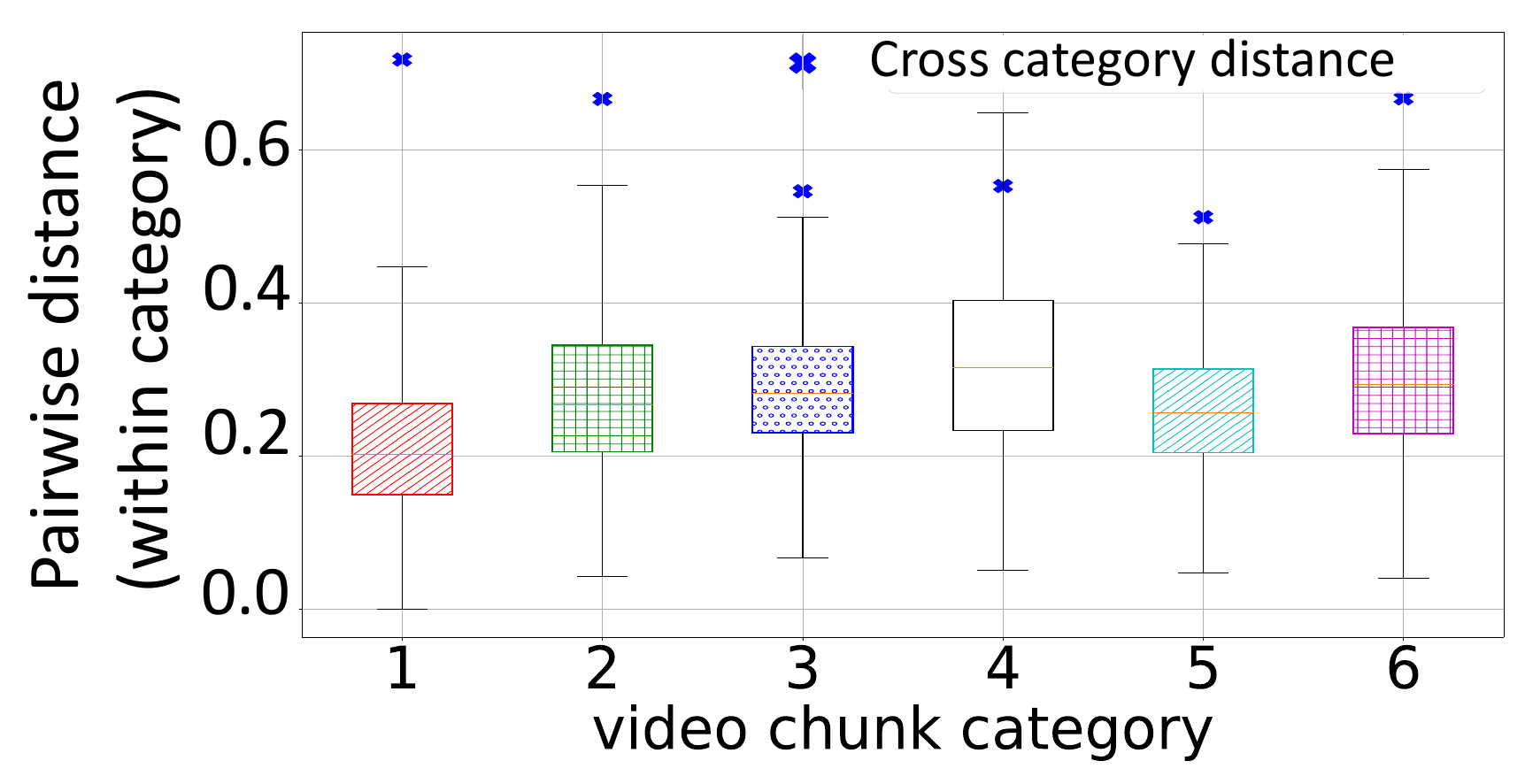}
    \label{fig:Similarity measurement within and between video categories New Metric}}
\vspace{-2mm}
\caption{(a)- DB-score analysis to determine the optimum number of categories, (b)- \textit{within-category} and \textit{cross-category} analysis of pairwise distance measurement calculated according to Eq.~\ref{eq:similarity_metric2} }\vspace{-4mm}
\label{fig:Optimal number of video categories}
\end{figure}

We conducted a Davies-Bouldin score analysis to determine the optimum number of categories as shown in Fig.~\ref{fig:DB-Score_analysis_for_optimum_Q}, which yielded \textit{Q}=6. 
Fig.~\ref{fig:overview_dynamic_video_cat} summarizes the feature extraction and categorization of video chunks. 
Fig.~\ref{fig:Similarity measurement within and between video categories New Metric} represent the \textit{within-category} and \textit{cross-category} evaluation of the six dynamic video chunk categories. On average, the within-cluster pairwise feature distance is 0.27 and the cross-cluster pairwise feature distance is 0.56. 


\vspace{-2mm}



Fig.~\ref{fig:Video categories heatmaps and feature hist} shows heatmaps of equirectangular spatial viewport distribution for all six dynamic video categories obtained by combining the viewport traces of all of the users corresponding to its video chunks. A clear difference is observed between different spatial viewport distributions. In Fig.~\ref{fig:Video categories heatmaps and feature hist}, a histogram representation of the feature vectors of video chunks in each category is indicated below each corresponding heatmap. It is evident that the proposed dynamic video categorization algorithm successfully identifies different combinations of user trace types for video chunk clustering. 

\textit{\textbf{Takeaway- }In viewing any \ang{360} video, there are different users who display different behaviors. It is advantageous to take all of these different behaviors into account, rather than generalizing it to the majority behavior when characterizing a video. The proposed video chunk categorization clusters the videos that induce similar types of behaviors in their users, irrespective of the video content.}


\vspace{-1mm}
\subsection{Static vs Dynamic video chunk clustering}

\begin{figure}[h!]
\centering
\subfloat[\# of dynamic clusters per video  ]{\includegraphics[width=0.22\textwidth]{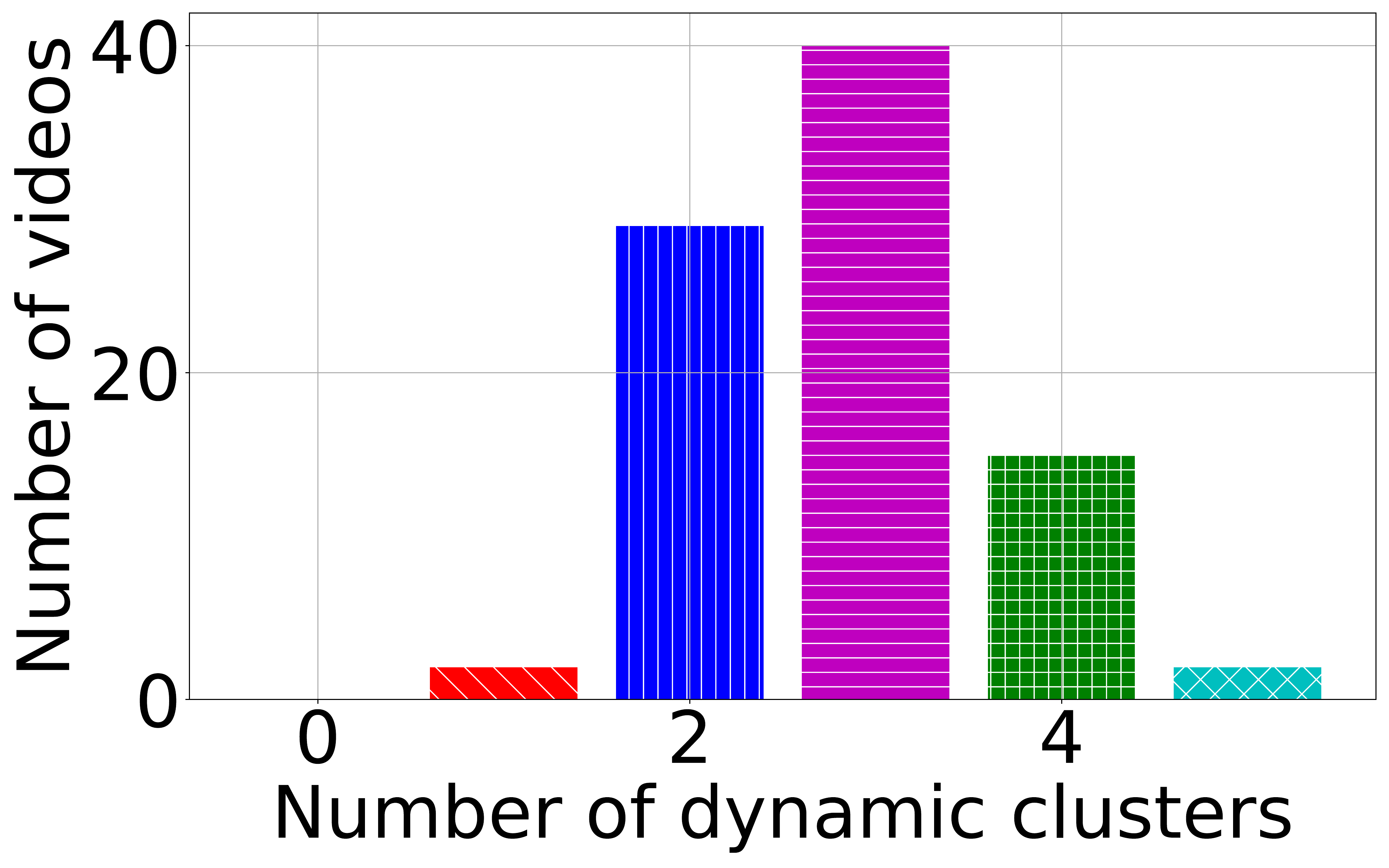}
    \label{fig:DynamicClustersPerVideo}}
    \hspace{0.5mm}
\subfloat[Static vs dynamic cat. ]{\includegraphics[width=0.22\textwidth]{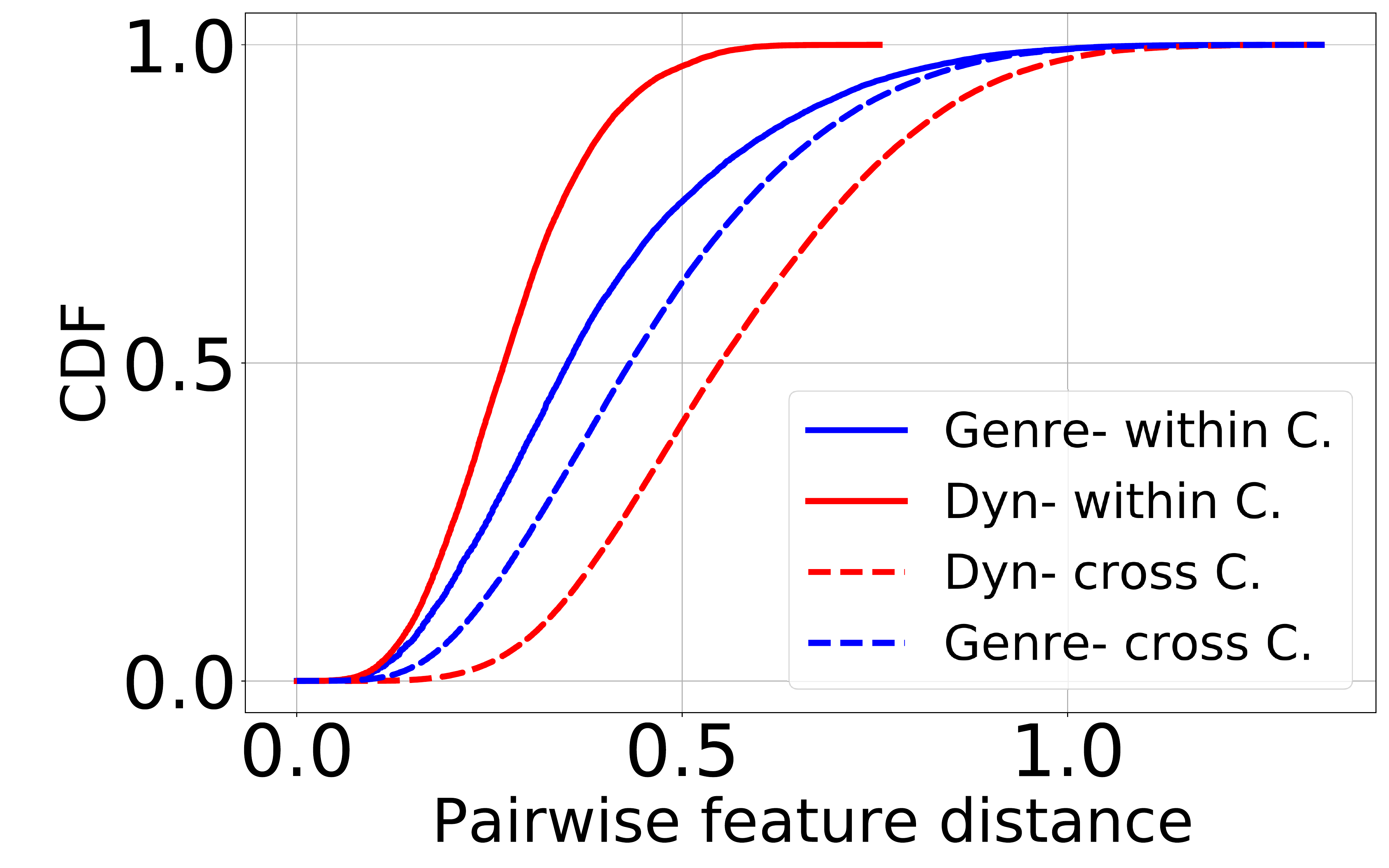}
    \label{fig:static vs dynamic new metric}}
    \vspace{-2mm}
\caption{a)- \# of dynamic clusters per a particular video b)- Comparison static vs dynamic categorizations using Eq \ref{eq:similarity_metric2}}\vspace{-2mm}
\label{fig:dynamic cat plots}
\end{figure}

Majority of the existing categorizations of \ang{360} videos are static; the entire video belongs to a particular category. These static methods often fail since significant temporal variations in video content and corresponding user behavior can occur in long videos.
To the best of our knowledge, ours is the first proposed algorithm that dynamically categorizes \ang{360} videos by taking 2s video chunks. Fig \ref{fig:DynamicClustersPerVideo} shows how different chunks of the same video can belong to different categories according to the proposed algorithm.

We compared the performance of the proposed categorization with genre-based static categorization by \cite{afzal_characterization_2017}. The 88 videos in the dataset were assigned to 10 genres such as \textit{Sports, scenety etc.}
Fig.~\ref{fig:static vs dynamic new metric} shows the comparison results using \textit{within-category} and \textit{cross-category} feature distances as introduced in Eq.~\ref{eq:similarity_metric2}. 


\vspace{-2mm}

\textit{\textbf{Takeaway- }Static categorizations of a long \ang{360} videos fail to take temporal variations of contents and user behaviors into account. Breaking the videos down to independant chunks enables more accurate categorizations. Quantifying the categorization in terms of user viewport patterns eliminates ambiguities and the requirement of semantic information for categorization.}

\vspace{-2mm}
\section{Conclusion and Future work}

In this study, we conducted an in-depth analysis of user behaviors in \ang{360} video streaming using a large dataset of over 3700 viewport traces. 
By extracting both spatial and temporal features from viewport traces, the proposed method can identify viewport similarities across different videos whilst achieving 
81.17\% \textit{within-cluster} viewport overlap and outperforms the existing viewport clustering algorithms.
We extended the above solution to develop a novel \ang{360} video categorization which successfully captures temporal changes of content and user behaviors by independently processing 2s video chunks. The proposed categorization shows 28.32\% improvement in clustering videos with similar combinations of user behaviors over existing genre-based \ang{360} video categorizations. 

In future work, we aim to extend our video categorization method for to develop a novel caching algorithm to enhance the \ang{360} video processing at edge servers. We plan to explore new approaches for edge resource optimization further analysing the content of the categorized videos.

\bibliographystyle{ACM-Reference-Format}
\newpage
\balance
\bibliography{sample-base}


\begin{thebibliography}{24}


\ifx \showCODEN    \undefined \def \showCODEN     #1{\unskip}     \fi
\ifx \showDOI      \undefined \def \showDOI       #1{#1}\fi
\ifx \showISBNx    \undefined \def \showISBNx     #1{\unskip}     \fi
\ifx \showISBNxiii \undefined \def \showISBNxiii  #1{\unskip}     \fi
\ifx \showISSN     \undefined \def \showISSN      #1{\unskip}     \fi
\ifx \showLCCN     \undefined \def \showLCCN      #1{\unskip}     \fi
\ifx \shownote     \undefined \def \shownote      #1{#1}          \fi
\ifx \showarticletitle \undefined \def \showarticletitle #1{#1}   \fi
\ifx \showURL      \undefined \def \showURL       {\relax}        \fi
\providecommand\bibfield[2]{#2}
\providecommand\bibinfo[2]{#2}
\providecommand\natexlab[1]{#1}
\providecommand\showeprint[2][]{arXiv:#2}

\bibitem[\protect\citeauthoryear{Afzal, Chen, and Ramakrishnan}{Afzal
  et~al\mbox{.}}{2017}]%
        {afzal_characterization_2017}
\bibfield{author}{\bibinfo{person}{Shahryar Afzal}, \bibinfo{person}{Jiasi
  Chen}, {and} \bibinfo{person}{K.~K. Ramakrishnan}.}
  \bibinfo{year}{2017}\natexlab{}.
\newblock \showarticletitle{Characterization of 360-degree {Videos}}. In
  \bibinfo{booktitle}{\emph{Proceedings of the {Workshop} on {Virtual}
  {Reality} and {Augmented} {Reality} {Network} - {VR}/{AR} {Network} '17}}.
  \bibinfo{publisher}{ACM Press}, \bibinfo{address}{Los Angeles, CA, USA},
  \bibinfo{pages}{1--6}.
\newblock
\showISBNx{978-1-4503-5055-6}
\urldef\tempurl%
\url{https://doi.org/10.1145/3097895.3097896}
\showDOI{\tempurl}


\bibitem[\protect\citeauthoryear{Almquist, Almquist, Krishnamoorthi, Carlsson,
  and Eager}{Almquist et~al\mbox{.}}{2018}]%
        {almquist_prefetch_2018}
\bibfield{author}{\bibinfo{person}{Mathias Almquist}, \bibinfo{person}{Viktor
  Almquist}, \bibinfo{person}{Vengatanathan Krishnamoorthi},
  \bibinfo{person}{Niklas Carlsson}, {and} \bibinfo{person}{Derek Eager}.}
  \bibinfo{year}{2018}\natexlab{}.
\newblock \showarticletitle{The {Prefetch} {Aggressiveness} {Tradeoff} in
  360\${\textasciicircum}\{{\textbackslash}circ\}\$ {Video} {Streaming}}.
\newblock \bibinfo{journal}{\emph{Proceedings of the 9th ACM Multimedia Systems
  Conference}} (\bibinfo{date}{June} \bibinfo{year}{2018}),
  \bibinfo{pages}{258--269}.
\newblock
\urldef\tempurl%
\url{https://doi.org/10.1145/3204949.3204970}
\showDOI{\tempurl}
\newblock
\shownote{arXiv: 1812.07277.}


\bibitem[\protect\citeauthoryear{Bao, Wu, Zhang, Ramli, and Liu}{Bao
  et~al\mbox{.}}{2016}]%
        {bao_shooting_2016}
\bibfield{author}{\bibinfo{person}{Yanan Bao}, \bibinfo{person}{Huasen Wu},
  \bibinfo{person}{Tianxiao Zhang}, \bibinfo{person}{Albara~Ah Ramli}, {and}
  \bibinfo{person}{Xin Liu}.} \bibinfo{year}{2016}\natexlab{}.
\newblock \showarticletitle{Shooting a moving target: {Motion}-prediction-based
  transmission for 360-degree videos}. In \bibinfo{booktitle}{\emph{2016 {IEEE}
  {International} {Conference} on {Big} {Data} ({Big} {Data})}}.
  \bibinfo{publisher}{IEEE}, \bibinfo{address}{Washington DC,USA},
  \bibinfo{pages}{1161--1170}.
\newblock
\showISBNx{978-1-4673-9005-7}
\urldef\tempurl%
\url{https://doi.org/10.1109/BigData.2016.7840720}
\showDOI{\tempurl}


\bibitem[\protect\citeauthoryear{Carlsson and Eager}{Carlsson and
  Eager}{2020}]%
        {carlsson_had_2020}
\bibfield{author}{\bibinfo{person}{Niklas Carlsson} {and}
  \bibinfo{person}{Derek Eager}.} \bibinfo{year}{2020}\natexlab{}.
\newblock \showarticletitle{Had {You} {Looked} {Where} {I}'m {Looking}?
  {Cross}-user {Similarities} in {Viewing} {Behavior} for 360° {Video} and
  {Caching} {Implications}}.
\newblock  (\bibinfo{year}{2020}), \bibinfo{pages}{8}.
\newblock


\bibitem[\protect\citeauthoryear{Corbillon, De~Simone, and Simon}{Corbillon
  et~al\mbox{.}}{2017}]%
        {corbillon_360-degree_2017}
\bibfield{author}{\bibinfo{person}{Xavier Corbillon},
  \bibinfo{person}{Francesca De~Simone}, {and} \bibinfo{person}{Gwendal
  Simon}.} \bibinfo{year}{2017}\natexlab{}.
\newblock \showarticletitle{360-{Degree} {Video} {Head} {Movement} {Dataset}}.
  In \bibinfo{booktitle}{\emph{Proceedings of the 8th {ACM} on {Multimedia}
  {Systems} {Conference}}}. \bibinfo{publisher}{ACM}, \bibinfo{address}{Taipei
  Taiwan}, \bibinfo{pages}{199--204}.
\newblock
\showISBNx{978-1-4503-5002-0}
\urldef\tempurl%
\url{https://doi.org/10.1145/3083187.3083215}
\showDOI{\tempurl}


\bibitem[\protect\citeauthoryear{Facebook}{Facebook}{[n.d.]a}]%
        {fb_360}
\bibfield{author}{\bibinfo{person}{Facebook}.}
  \bibinfo{year}{[n.d.]}\natexlab{a}.
\newblock \bibinfo{booktitle}{\emph{Facebook360}}.
\newblock
\urldef\tempurl%
\url{https://facebook360.fb.com/live360/}
\showURL{%
\tempurl}


\bibitem[\protect\citeauthoryear{Facebook}{Facebook}{[n.d.]b}]%
        {fb_oculus}
\bibfield{author}{\bibinfo{person}{Facebook}.}
  \bibinfo{year}{[n.d.]}\natexlab{b}.
\newblock \bibinfo{booktitle}{\emph{Oculus From Facebook}}.
\newblock
\urldef\tempurl%
\url{https://www.oculus.com}
\showURL{%
\tempurl}


\bibitem[\protect\citeauthoryear{Guan, Zheng, Zhang, Guo, and Jiang}{Guan
  et~al\mbox{.}}{2019}]%
        {guan_pano_2019}
\bibfield{author}{\bibinfo{person}{Yu Guan}, \bibinfo{person}{Chengyuan Zheng},
  \bibinfo{person}{Xinggong Zhang}, \bibinfo{person}{Zongming Guo}, {and}
  \bibinfo{person}{Junchen Jiang}.} \bibinfo{year}{2019}\natexlab{}.
\newblock \showarticletitle{Pano: optimizing 360° video streaming with a
  better understanding of quality perception}. In
  \bibinfo{booktitle}{\emph{Proceedings of the {ACM} {Special} {Interest}
  {Group} on {Data} {Communication}}}. \bibinfo{publisher}{ACM},
  \bibinfo{address}{Beijing China}, \bibinfo{pages}{394--407}.
\newblock
\showISBNx{978-1-4503-5956-6}
\urldef\tempurl%
\url{https://doi.org/10.1145/3341302.3342063}
\showDOI{\tempurl}


\bibitem[\protect\citeauthoryear{He, Qureshi, Qiu, Li, Li, and Han}{He
  et~al\mbox{.}}{2018}]%
        {he2018rubiks}
\bibfield{author}{\bibinfo{person}{Jian He}, \bibinfo{person}{Mubashir~Adnan
  Qureshi}, \bibinfo{person}{Lili Qiu}, \bibinfo{person}{Jin Li},
  \bibinfo{person}{Feng Li}, {and} \bibinfo{person}{Lei Han}.}
  \bibinfo{year}{2018}\natexlab{}.
\newblock \showarticletitle{Rubiks: Practical 360-degree streaming for
  smartphones}. In \bibinfo{booktitle}{\emph{Proceedings of the 16th Annual
  International Conference on Mobile Systems, Applications, and Services}}.
  \bibinfo{pages}{482--494}.
\newblock


\bibitem[\protect\citeauthoryear{Jiang, Wu, Wang, Xue, and Chang}{Jiang
  et~al\mbox{.}}{2017}]%
        {jiang2017exploiting}
\bibfield{author}{\bibinfo{person}{Yu-Gang Jiang}, \bibinfo{person}{Zuxuan Wu},
  \bibinfo{person}{Jun Wang}, \bibinfo{person}{Xiangyang Xue}, {and}
  \bibinfo{person}{Shih-Fu Chang}.} \bibinfo{year}{2017}\natexlab{}.
\newblock \showarticletitle{Exploiting feature and class relationships in video
  categorization with regularized deep neural networks}.
\newblock \bibinfo{journal}{\emph{IEEE transactions on pattern analysis and
  machine intelligence}} \bibinfo{volume}{40}, \bibinfo{number}{2}
  (\bibinfo{year}{2017}), \bibinfo{pages}{352--364}.
\newblock


\bibitem[\protect\citeauthoryear{Lo, Fan, Lee, Huang, Chen, and Hsu}{Lo
  et~al\mbox{.}}{2017}]%
        {lo_360_2017}
\bibfield{author}{\bibinfo{person}{Wen-Chih Lo}, \bibinfo{person}{Ching-Ling
  Fan}, \bibinfo{person}{Jean Lee}, \bibinfo{person}{Chun-Ying Huang},
  \bibinfo{person}{Kuan-Ta Chen}, {and} \bibinfo{person}{Cheng-Hsin Hsu}.}
  \bibinfo{year}{2017}\natexlab{}.
\newblock \showarticletitle{360° {Video} {Viewing} {Dataset} in
  {Head}-{Mounted} {Virtual} {Reality}}. In
  \bibinfo{booktitle}{\emph{Proceedings of the 8th {ACM} on {Multimedia}
  {Systems} {Conference}}}. \bibinfo{publisher}{ACM}, \bibinfo{address}{Taipei
  Taiwan}, \bibinfo{pages}{211--216}.
\newblock
\showISBNx{978-1-4503-5002-0}
\urldef\tempurl%
\url{https://doi.org/10.1145/3083187.3083219}
\showDOI{\tempurl}


\bibitem[\protect\citeauthoryear{McClanahan and Gokhale}{McClanahan and
  Gokhale}{2017}]%
        {mcclanahan2017interplay}
\bibfield{author}{\bibinfo{person}{Brian McClanahan} {and}
  \bibinfo{person}{Swapna~S Gokhale}.} \bibinfo{year}{2017}\natexlab{}.
\newblock \showarticletitle{Interplay between video recommendations,
  categories, and popularity on YouTube}. In \bibinfo{booktitle}{\emph{2017
  IEEE SmartWorld, Ubiquitous Intelligence \& Computing, Advanced \& Trusted
  Computed, Scalable Computing \& Communications, Cloud \& Big Data Computing,
  Internet of People and Smart City Innovation
  (SmartWorld/SCALCOM/UIC/ATC/CBDCom/IOP/SCI)}}. IEEE, \bibinfo{pages}{1--7}.
\newblock


\bibitem[\protect\citeauthoryear{Microsoft}{Microsoft}{[n.d.]}]%
        {MS_hololens}
\bibfield{author}{\bibinfo{person}{Microsoft}.}
  \bibinfo{year}{[n.d.]}\natexlab{}.
\newblock \bibinfo{booktitle}{\emph{Microsoft Hololens}}.
\newblock
\urldef\tempurl%
\url{https://www.microsoft.com/en-us/hololens}
\showURL{%
\tempurl}


\bibitem[\protect\citeauthoryear{Nasrabadi, Samiei, Mahzari, McMahan, Prakash,
  Farias, and Carvalho}{Nasrabadi et~al\mbox{.}}{2019}]%
        {nasrabadi_taxonomy_2019}
\bibfield{author}{\bibinfo{person}{Afshin~Taghavi Nasrabadi},
  \bibinfo{person}{Aliehsan Samiei}, \bibinfo{person}{Anahita Mahzari},
  \bibinfo{person}{Ryan~P. McMahan}, \bibinfo{person}{Ravi Prakash},
  \bibinfo{person}{Mylène C.~Q. Farias}, {and} \bibinfo{person}{Marcelo~M.
  Carvalho}.} \bibinfo{year}{2019}\natexlab{}.
\newblock \showarticletitle{A taxonomy and dataset for 360° videos}. In
  \bibinfo{booktitle}{\emph{Proceedings of the 10th {ACM} {Multimedia}
  {Systems} {Conference}}}. \bibinfo{publisher}{ACM}, \bibinfo{address}{Amherst
  Massachusetts}, \bibinfo{pages}{273--278}.
\newblock
\showISBNx{978-1-4503-6297-9}
\urldef\tempurl%
\url{https://doi.org/10.1145/3304109.3325812}
\showDOI{\tempurl}


\bibitem[\protect\citeauthoryear{Petrangeli, Simon, and Swaminathan}{Petrangeli
  et~al\mbox{.}}{2018}]%
        {petrangeli_trajectory-based_2018}
\bibfield{author}{\bibinfo{person}{Stefano Petrangeli},
  \bibinfo{person}{Gwendal Simon}, {and} \bibinfo{person}{Viswanathan
  Swaminathan}.} \bibinfo{year}{2018}\natexlab{}.
\newblock \showarticletitle{Trajectory-{Based} {Viewport} {Prediction} for
  360-{Degree} {Virtual} {Reality} {Videos}}. In \bibinfo{booktitle}{\emph{2018
  {IEEE} {International} {Conference} on {Artificial} {Intelligence} and
  {Virtual} {Reality} ({AIVR})}}. \bibinfo{publisher}{IEEE},
  \bibinfo{address}{Taichung, Taiwan}, \bibinfo{pages}{157--160}.
\newblock
\showISBNx{978-1-5386-9269-1}
\urldef\tempurl%
\url{https://doi.org/10.1109/AIVR.2018.00033}
\showDOI{\tempurl}


\bibitem[\protect\citeauthoryear{Qian, Han, Xiao, and Gopalakrishnan}{Qian
  et~al\mbox{.}}{2018}]%
        {qian2018flare}
\bibfield{author}{\bibinfo{person}{Feng Qian}, \bibinfo{person}{Bo Han},
  \bibinfo{person}{Qingyang Xiao}, {and} \bibinfo{person}{Vijay
  Gopalakrishnan}.} \bibinfo{year}{2018}\natexlab{}.
\newblock \showarticletitle{Flare: Practical viewport-adaptive 360-degree video
  streaming for mobile devices}. In \bibinfo{booktitle}{\emph{Proceedings of
  the 24th Annual International Conference on Mobile Computing and
  Networking}}. \bibinfo{pages}{99--114}.
\newblock


\bibitem[\protect\citeauthoryear{Qian, Ji, Han, and Gopalakrishnan}{Qian
  et~al\mbox{.}}{2016}]%
        {qian_optimizing_2016}
\bibfield{author}{\bibinfo{person}{Feng Qian}, \bibinfo{person}{Lusheng Ji},
  \bibinfo{person}{Bo Han}, {and} \bibinfo{person}{Vijay Gopalakrishnan}.}
  \bibinfo{year}{2016}\natexlab{}.
\newblock \showarticletitle{Optimizing 360 video delivery over cellular
  networks}. In \bibinfo{booktitle}{\emph{Proceedings of the 5th {Workshop} on
  {All} {Things} {Cellular} {Operations}, {Applications} and {Challenges} -
  {ATC} '16}}. \bibinfo{publisher}{ACM Press}, \bibinfo{address}{New York City,
  New York}, \bibinfo{pages}{1--6}.
\newblock
\showISBNx{978-1-4503-4249-0}
\urldef\tempurl%
\url{https://doi.org/10.1145/2980055.2980056}
\showDOI{\tempurl}


\bibitem[\protect\citeauthoryear{Rossi, Simone, Frossard, and Toni}{Rossi
  et~al\mbox{.}}{2019}]%
        {rossi_spherical_2019}
\bibfield{author}{\bibinfo{person}{S. Rossi}, \bibinfo{person}{F.~De Simone},
  \bibinfo{person}{P. Frossard}, {and} \bibinfo{person}{L. Toni}.}
  \bibinfo{year}{2019}\natexlab{}.
\newblock \showarticletitle{Spherical {Clustering} of {Users} {Navigating}
  360° {Content}}. In \bibinfo{booktitle}{\emph{{ICASSP} 2019 - 2019 {IEEE}
  {International} {Conference} on {Acoustics}, {Speech} and {Signal}
  {Processing} ({ICASSP})}}. \bibinfo{pages}{4020--4024}.
\newblock
\urldef\tempurl%
\url{https://doi.org/10.1109/ICASSP.2019.8683854}
\showDOI{\tempurl}
\newblock
\shownote{ISSN: 2379-190X.}


\bibitem[\protect\citeauthoryear{Samsung}{Samsung}{[n.d.]}]%
        {samsung_gear}
\bibfield{author}{\bibinfo{person}{Samsung}.}
  \bibinfo{year}{[n.d.]}\natexlab{}.
\newblock \bibinfo{booktitle}{\emph{Samsung Gear VR}}.
\newblock
\urldef\tempurl%
\url{https://www.samsung.com/global/galaxy/gear-vr/}
\showURL{%
\tempurl}


\bibitem[\protect\citeauthoryear{Sun, Wu, Wang, Arai, Kinebuchi, and Jiang}{Sun
  et~al\mbox{.}}{2016}]%
        {sun2016exploiting}
\bibfield{author}{\bibinfo{person}{Yongqing Sun}, \bibinfo{person}{Zuxuan Wu},
  \bibinfo{person}{Xi Wang}, \bibinfo{person}{Hiroyuki Arai},
  \bibinfo{person}{Tetsuya Kinebuchi}, {and} \bibinfo{person}{Yu-Gang Jiang}.}
  \bibinfo{year}{2016}\natexlab{}.
\newblock \showarticletitle{Exploiting objects with LSTMs for video
  categorization}. In \bibinfo{booktitle}{\emph{Proceedings of the 24th ACM
  international conference on Multimedia}}. \bibinfo{pages}{142--146}.
\newblock


\bibitem[\protect\citeauthoryear{Wu, Tan, Wang, and Yang}{Wu
  et~al\mbox{.}}{2017}]%
        {wu_dataset_2017}
\bibfield{author}{\bibinfo{person}{Chenglei Wu}, \bibinfo{person}{Zhihao Tan},
  \bibinfo{person}{Zhi Wang}, {and} \bibinfo{person}{Shiqiang Yang}.}
  \bibinfo{year}{2017}\natexlab{}.
\newblock \showarticletitle{A {Dataset} for {Exploring} {User} {Behaviors} in
  {VR} {Spherical} {Video} {Streaming}}. In
  \bibinfo{booktitle}{\emph{Proceedings of the 8th {ACM} on {Multimedia}
  {Systems} {Conference}}}. \bibinfo{publisher}{ACM}, \bibinfo{address}{Taipei
  Taiwan}, \bibinfo{pages}{193--198}.
\newblock
\showISBNx{978-1-4503-5002-0}
\urldef\tempurl%
\url{https://doi.org/10.1145/3083187.3083210}
\showDOI{\tempurl}


\bibitem[\protect\citeauthoryear{Xie, Zhang, and Guo}{Xie
  et~al\mbox{.}}{2018}]%
        {xie_cls_2018}
\bibfield{author}{\bibinfo{person}{Lan Xie}, \bibinfo{person}{Xinggong Zhang},
  {and} \bibinfo{person}{Zongming Guo}.} \bibinfo{year}{2018}\natexlab{}.
\newblock \showarticletitle{{CLS}: {A} {Cross}-user {Learning} based {System}
  for {Improving} {QoE} in 360-degree {Video} {Adaptive} {Streaming}}. In
  \bibinfo{booktitle}{\emph{2018 {ACM} {Multimedia} {Conference} on
  {Multimedia} {Conference} - {MM} '18}}. \bibinfo{publisher}{ACM Press},
  \bibinfo{address}{Seoul, Republic of Korea}, \bibinfo{pages}{564--572}.
\newblock
\showISBNx{978-1-4503-5665-7}
\urldef\tempurl%
\url{https://doi.org/10.1145/3240508.3240556}
\showDOI{\tempurl}


\bibitem[\protect\citeauthoryear{Xu, Han, and Qian}{Xu et~al\mbox{.}}{2019}]%
        {xu_analyzing_2019}
\bibfield{author}{\bibinfo{person}{Tan Xu}, \bibinfo{person}{Bo Han}, {and}
  \bibinfo{person}{Feng Qian}.} \bibinfo{year}{2019}\natexlab{}.
\newblock \showarticletitle{Analyzing viewport prediction under different {VR}
  interactions}. In \bibinfo{booktitle}{\emph{Proceedings of the 15th
  {International} {Conference} on {Emerging} {Networking} {Experiments} {And}
  {Technologies}}}. \bibinfo{publisher}{ACM}, \bibinfo{address}{Orlando
  Florida}, \bibinfo{pages}{165--171}.
\newblock
\showISBNx{978-1-4503-6998-5}
\urldef\tempurl%
\url{https://doi.org/10.1145/3359989.3365413}
\showDOI{\tempurl}


\bibitem[\protect\citeauthoryear{{YT Creater Acedemy}}{{YT Creater
  Acedemy}}{[n.d.]}]%
        {yt_360}
\bibfield{author}{\bibinfo{person}{{YT Creater Acedemy}}.}
  \bibinfo{year}{[n.d.]}\natexlab{}.
\newblock \bibinfo{booktitle}{\emph{Introduction to 360-degree video and
  virtual reality}}.
\newblock
\urldef\tempurl%
\url{https://creatoracademy.youtube.com/page/lesson/spherical-video}
\showURL{%
\tempurl}


\end{thebibliography}

\end{document}